\documentclass[10pt,a4paper]{article}

\usepackage{report_style}

\hypersetup{pdftitle={Do Alice and Bob Dream of the Oracle? Wishful Cryptanalysis},pdfauthor={Guoxiao Liu, Keting Jia, Hongbo Yu, Zongyue Wang, Tianshuo Cong, Junfeng Fan, Anyu Wang, Xiaoyun Wang}}

\begin{document}

\thispagestyle{firstpage}

\vspace*{-20pt}

\reportherocard{%
{\sffamily\fontsize{15}{18.5}\selectfont\bfseries\color{ReportPrimary}\noindent\mbox{Do Alice and Bob Dream of the Oracle? Wishful Cryptanalysis}\par}
\vspace{5pt}
\noindent\begin{minipage}[c]{.79\linewidth}\raggedright
{\sffamily\fontsize{8.8}{11.8}\selectfont\bfseries \mbox{Guoxiao Liu\textsuperscript{1}}, \mbox{Keting Jia\textsuperscript{1}}, \mbox{Hongbo Yu\textsuperscript{1}}, \mbox{Zongyue Wang\textsuperscript{2}}, \mbox{Tianshuo Cong\textsuperscript{3}}, \mbox{Junfeng Fan\textsuperscript{2}}, \mbox{Anyu Wang\textsuperscript{1}}, \mbox{Xiaoyun Wang\textsuperscript{1\,\corrmail}}\par}
\vspace{2.5pt}
{\sffamily\fontsize{7.3}{9.8}\selectfont\color{ReportGray}\textsuperscript{1}Tsinghua University\hspace{0.55em}\textsuperscript{2}Open Security Research\hspace{0.55em}\textsuperscript{3}Shandong University\hspace{0.7em}\mbox{\corrmail\hspace{0.22em}Corresponding author}\par}
\end{minipage}\hfill\begin{minipage}[c]{.195\linewidth}\raggedleft
\includegraphics[width=\linewidth,height=58pt,keepaspectratio]{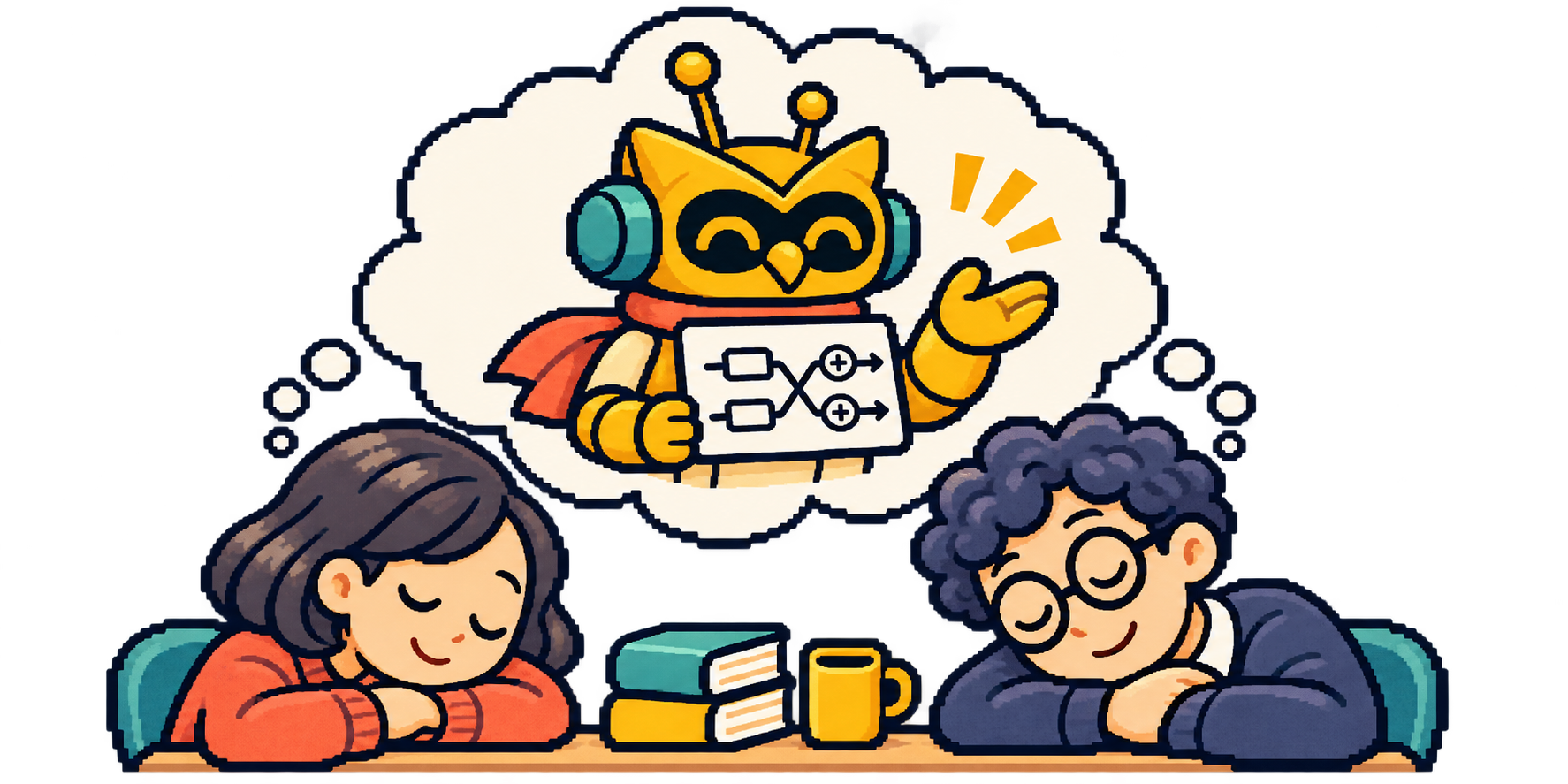}
\end{minipage}\par
\vspace{5.5pt}{\color{ReportRule}\hrule height 0.4pt}\vspace{4.5pt}
{\centering\fontsize{8.8}{11.6}\selectfont\color{ReportGray}{\itshape ``The only way of discovering the limits of the possible is to venture a little way past them into the impossible.''}\enspace\mbox{\upshape---Arthur C. Clarke}\par}
\vspace{4.5pt}{\color{ReportRule}\hrule height 0.4pt}\vspace{5.5pt}
{\sffamily\fontsize{7.8}{9.8}\selectfont\bfseries\color{ReportPrimary}ABSTRACT}\par\vspace{2.5pt}
{\fontsize{9.1}{12.0}\selectfont\noindent For two millennia, the history of cryptography has been one of security boundaries forged through offense and defense. Tools steadily automated calculation and search, yet human minds remained the sole protagonists of insight and discovery. Foundation models now mark an inflection point from automated computation toward intelligent scientific discovery. Because cryptanalysis rests on exact mathematical definitions and immediately executable truth, it offers a natural proving ground for autonomous intelligence to move past security reliant on scarce human attention and explore structures beyond human working memory. Yet intelligent discovery does not arrive merely by wishing for answers. Unguided models stall between two opposite traps: \textit{authority hallucination}, where they surrender whenever prior literature declares a route blocked, and \textit{reasoning hallucination}, where they produce subtle, unverifiable flaws. Closed platforms and generic agent loops further impede sustained open inquiry. We introduce \textbf{wishful cryptanalysis}, a harness-agnostic methodology that pairs constructive optimism against entrenched assumptions with blinded, zero-import verification against error. Agents first compress reduced-scale theory within provisioned compute so live witnesses certify their mathematics automatically. They then co-adapt primitive, algorithm, and hardware, and climb parameter ladders toward high-round theory. Applied across frontier open-weight and proprietary models with minimal human intervention beyond targets and compute, this paradigm yielded novel attack techniques and thirteen verifiable practical results across Keccak (SHA-3) and AES. These include breaking a fifteen-year-open public challenge and closing longstanding practical gaps, released alongside an early-preview open-source workbench (\texttt{agent-\allowbreak{}framework}). Extending Kerckhoffs's principle into the AI era, we envision an \textbf{Open Oracle}, an open and auditable public infrastructure that frees cryptographers to legislate security goals, challenge assumptions, distill theory, and co-design resilient standards.\par}
\vspace{6pt}{\color{ReportRule}\hrule height 0.4pt}\vspace{5pt}
{\sffamily\fontsize{7.6}{10.5}\selectfont\color{ReportGray}\noindent
\textbf{\color{ReportPrimary}Date:}\enspace November 20, 2026 (Preview)\par\vspace{1.2pt}
\textbf{\color{ReportPrimary}Project and code:}\enspace\url{https://github.com/DayanCryptoLab/WishfulCryptanalysis}\par\vspace{1.2pt}
\textbf{\color{ReportPrimary}Correspondence:}\enspace \href{mailto:lgx22@mails.tsinghua.edu.cn}{\texttt{lgx22@mails.tsinghua.edu.cn}}, \href{mailto:xiaoyunwang@tsinghua.edu.cn}{\texttt{xiaoyunwang@tsinghua.edu.cn}}\par\vspace{1.2pt}
\textbf{\color{ReportPrimary}Keywords:}\enspace Wishful cryptanalysis $\cdot$ Open oracle $\cdot$ Multi-agent systems $\cdot$ Executable verification $\cdot$ Cryptographic algorithms\par}
}

\vspace{5pt}

\begin{center}
\vspace{0pt}
\begin{tikzpicture}
\node[anchor=south west,inner sep=0] (img) at (0,0) {\includegraphics[width=.92\linewidth,height=.162\textheight,keepaspectratio]{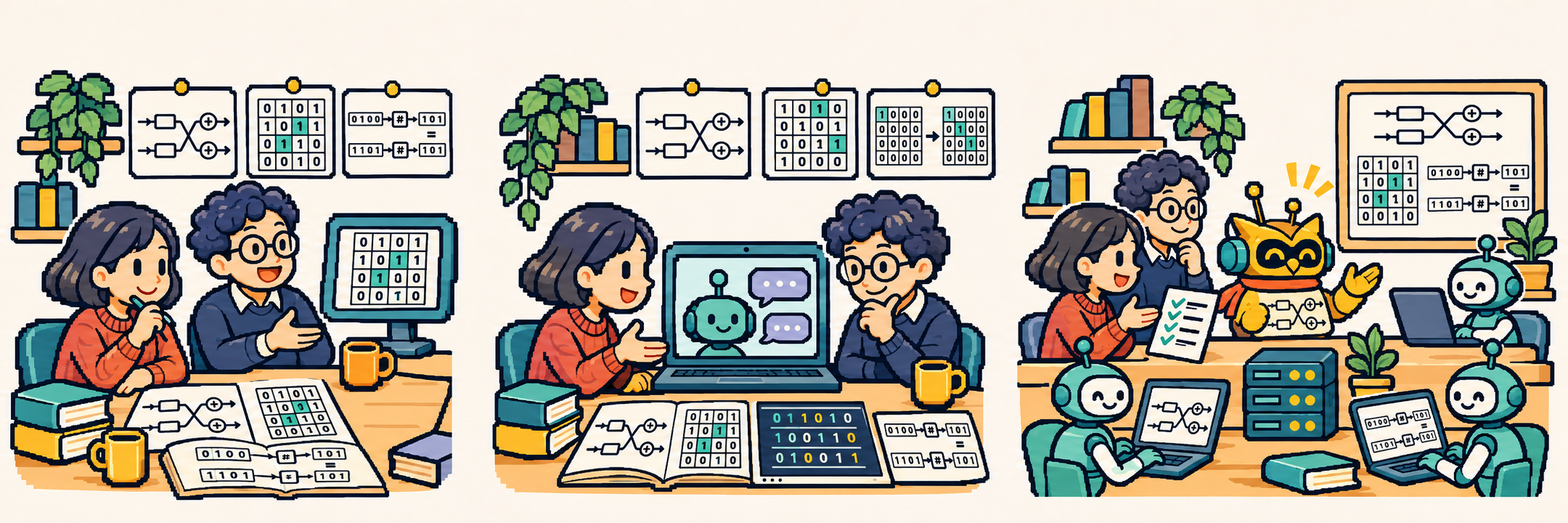}};
\begin{scope}[x={(img.south east)},y={(img.north west)}]
\node[anchor=north,inner sep=0pt,font=\footnotesize\sffamily\bfseries,text=ReportPrimary] at (0.1492,-0.045) {Human-led};
\node[anchor=north,inner sep=0pt,font=\footnotesize\sffamily\bfseries,text=ReportPrimary] at (0.4724,-0.045) {AI-assisted};
\node[anchor=north,inner sep=0pt,font=\footnotesize\sffamily\bfseries,text=ReportPrimary] at (0.8202,-0.045) {Agent-led};
\end{scope}
\end{tikzpicture}
\captionsetup{skip=3.5pt}\captionof{figure}{Three coexisting eras of cryptanalysis: \textbf{Human-led} derivation and automated search, \textbf{AI-assisted} cryptographic research, and \textbf{Agent-led} 24/7 autonomous discovery where human cryptographers define security goals, challenge hidden assumptions, and distill mathematical breakthroughs.}
\label{fig:paradigm}
\end{center}

\clearpage

{\small\setlength{\parskip}{1.6pt}\tableofcontents}

\clearpage

\Needspace{10\baselineskip}

\section{\texorpdfstring{Introduction: From Alice and Bob to the Research Oracle}{Introduction: From Alice and Bob to the Research Oracle}}\label{sec:intro}

The two-millennium history of cryptography is the history of security boundaries continually reforged through offense and defense. Security is never bestowed by design alone. It is forged and calibrated under relentless open analysis. From Caesar's shift cipher and the Vigen\`ere cipher, celebrated for three centuries as unbreakable, to the wartime cryptanalysis of Enigma at Bletchley Park that helped midwife the modern digital computer, every fallen wall has given birth to deeper cryptographic insight. In the modern era, Kerckhoffs's principle of open scrutiny \citep{kerckhoffs} elevated this duel into a rigorous mathematical science in which cryptanalysis charts and raises algorithmic security boundaries. Scrutinizing DES \citep{des} spurred differential cryptanalysis \citep{biham91} and quantifiable resistance criteria, compelling the global AES competition \citep{aes} to establish a new boundary anchored in the wide-trail strategy. RSA \citep{rsa} introduced Alice and Bob and grounded public-key security in explicit computational hardness assumptions, turning offense and defense into a continuous survey of mathematical intractability. Later, the practical collision breaks of MD5 by Xiaoyun Wang et al. \citep{mdfive,wang} shattered complacency around a ubiquitous standard and drove the field to build a new sponge-based security boundary in SHA-3 (Keccak) \citep{sha3}. Cryptography was also among the earliest disciplines to embrace automation, advancing from pen and paper to electromechanical Bombes, supercomputers, and SAT/MILP solvers. Yet however far automation progressed, machines supplied only automated computation and search within human-specified rules. Human minds remained the sole locus of intelligent cognition, framing the questions, spotting hidden symmetries, dismantling ``unbreakable'' orthodoxies, and redrawing the boundaries of security.

The emergence of foundation models first ushered in a transitional era of \textit{AI-assisted cryptographic research} \citep{chatgpt}. Researchers began consulting models to digest unfamiliar literature, check local derivations, or draft exploratory scripts, while still formulating hypotheses and orchestrating the entire inquiry themselves. Today, a far deeper shift is underway toward \textit{agent-led} investigation (Figure \ref{fig:paradigm}). Models now inhabit the computational environment twenty-four hours a day, searching literature, pitting competing attack hypotheses against one another, driving algebraic and combinatorial solvers, and refining multi-node SIMD and GPU kernels \citep{anthropic,ngcc}. For the first time in two thousand years, machines are crossing the watershed between automated search and intelligent discovery, stepping beyond obedient calculation directly into the reasoning arena of cryptanalysis. The \textbf{Oracle}, long an idealized abstraction in theoretical security proofs, is beginning to take shape in reality. Yet the leap from automated computation to intelligent scientific discovery is far from self-guaranteeing. Once machines become autonomous explorers capable of bold conjectures, yet equally susceptible to inherited orthodoxy and subtle error, a foundational question demands a systematic answer: \textit{what open, rigorous, self-certifying methodology can turn emerging machine intelligence into a trustworthy engine of cryptanalytic discovery?}

\Needspace{6\baselineskip}

\subsection{\texorpdfstring{Opportunities and challenges in the era of foundation models}{Opportunities and challenges in the era of foundation models}}\label{sec:opportunities}

Among all mathematical and engineering sciences, cryptanalysis is uniquely suited to autonomous intelligence. Unlike domains hampered by ambiguous evaluation metrics or slow wet-lab experiments, cryptanalysis rests on mathematically complete security models paired with immediate, indisputable executable truth. Its tasks span secret recovery of keys, plaintexts, or private parameters, distinguishing advantage, forgery and collision finding, underlying algebraic, lattice, and coding hardness instances, and cryptanalytic model extraction. Whether a candidate key decrypts, a witness passes specification checks, or a distinguisher or extraction attack meets its quantitative criterion can be verified directly by code. Precisely because its objectives are exact and its feedback is instantaneous, cryptanalysis offers an ideal proving ground for moving from automated computation to intelligent discovery. Anchored in a reliable methodology, autonomous agents unlock three progressively deeper opportunities for the field.

\begin{enumerate}

\item \textbf{Ending security by scarcity of human attention.} Specialist attention has always been finite. For decades, the sharpest human analysts concentrated on a few flagship standards such as AES and SHA-3, leaving scores of lightweight ciphers, post-quantum candidates, zero-knowledge-friendly hashes, and domain protocols merely under-analyzed. Tireless 24/7 agent teams make deep, uniform scrutiny scalable across every primitive before deployment.

\item \textbf{Collapsing the theory-to-silicon verification loop.} Hand-coding a multi-stage algebraic connector and SIMD/GPU search engine once took weeks, forcing academia to rely heavily on unverified high-round paper derivations. Agents close the loop from symbolic derivation to hardware execution in hours, making it practical to certify LLM theory through live reduced-round witnesses before scaling upward.

\item \textbf{Transcending cognitive silos toward new attack methods and paradigms.} Human researchers can hold only a handful of trail constraints, degree-drop identities, and hardware bottlenecks in working memory at once, and rarely have time to re-test ten discarded parameterizations of an old paper across differential, algebraic, and coding-theoretic silos. Agents already move beyond routine recombination to formulate novel attack techniques and structural inversions. Over the longer horizon, they point toward transcending human cognitive limits to invent entirely new cryptanalytic paradigms.

\end{enumerate}

Yet crossing from automated computation to intelligent discovery does not happen simply by making a wish to a foundation model. Without purpose-built epistemic and architectural mechanisms, invoking raw LLMs stalls on nontrivial cryptanalysis across three core barriers.

\begin{enumerate}

\item \textbf{Dual epistemic hallucinations.} Left unguided, models drift toward two opposite failure modes. At one extreme lies \textit{authority hallucination}, where a model surrenders prematurely whenever prior literature labels a route ``blocked,'' mistaking an obstacle of one restricted ansatz for cipher-wide impossibility much as generations once presumed Vigen\`ere or MD5 secure. At the opposite extreme lies \textit{reasoning hallucination}, where models miscount degrees of freedom, omit connector gates, or drift outside legal parameter windows in unverified high-round paper proofs that overwhelm manual human refereeing.

\item \textbf{Fragmented long-horizon exploration and the paper-to-hardware gap.} Deep cryptanalytic breakthroughs rarely emerge along a single linear chain of thought. They arise from splicing complementary partial insights across parallel threads, combining front-round linearizations, middle-round invariants, tail subspaces, and decisive counterexamples, and grounding those ideas in real hardware execution rather than nominal paper step counts. Generic single-loop coding harnesses and flat group chats lack both blinded adversarial isolation and structured mechanisms for synthesizing partial wins.

\item \textbf{Closed infrastructure and friction in frontier security research.} Sustained, auditable cryptanalysis requires tight integration with local solvers, compute clusters, and reproducible checkpoints. Meanwhile, when legitimate academic research pushes the boundaries of cryptanalytic security, closed proprietary platforms built around general-purpose guardrails frequently trigger over-cautious refusals, while black-box access undermines open scientific reproducibility.

\end{enumerate}

\Needspace{6\baselineskip}

\subsection{\texorpdfstring{Open models and the case for an Open Oracle}{Open models and the case for an Open Oracle}}\label{sec:open_oracle}

The third barrier raises a question deeper than engineering convenience: who should own the analytical intelligence that evaluates cryptographic security? In 1883, Auguste Kerckhoffs established that trust in a cipher must never rest on secrecy \citep{kerckhoffs}. Carrying that principle into the foundation-model era means building analytical intelligence itself as open, transparent, and auditable public infrastructure, an \textbf{Open Oracle}, rather than confining it inside institutional black boxes. Openness is equally vital to offset offense--defense asymmetry. Designing and standardizing a new cipher takes five to ten years, whereas automated structural breaks replicate at the speed of compute. Only when defenders, cipher designers, and public evaluators share equal analytical power can security boundaries be charted systematically before deployment. Open-weight reasoning and agentic models, including the DeepSeek \citep{deepseekr1}, GLM \citep{glm45}, and Kimi \citep{kimik2} families, together with research-friendly interfaces now supply a practical foundation for that oracle. Deployed locally or on institutional clusters, they couple directly with algebraic and combinatorial solvers under pinned checkpoints while sparing legitimate boundary research from the false-positive safety over-refusals that frequently abort long derivations on closed commercial platforms.

An open model substrate is necessary, yet not sufficient. Dropping an open model into a generic coding harness still runs headlong into the dual hallucinations and long-horizon synthesis bottlenecks of Section \ref{sec:opportunities}. Realizing intelligent discovery therefore requires a research methodology engineered around how foundation models reason, err, and collaborate. To that end, this report introduces \textbf{wishful cryptanalysis}, a harness-agnostic methodology built to surmount all three barriers and formalized in an open-source multi-agent workbench, alongside thirteen verified practical results (twenty-five self-certifying witnesses, summarized in Table \ref{tab:keccakresults}) across Keccak sponges and all five standard FIPS 197 \texttt{AES-128} hashing tracks \citep{sha3,crunchy,aes}.\footnote{Standalone witness checkers, theoretical complexity bounds, machine-checked \texttt{UNSAT} certificates, and the early-preview \texttt{agent-\allowbreak{}framework} (Section \ref{sec:ecosystem}) are archived at \url{https://github.com/DayanCryptoLab/WishfulCryptanalysis}.} Total LLM inference costs for these thirteen results were under US\$1,000, excluding downstream execution compute (Section \ref{sec:rigor_flywheel}). To make the principle of executable truth concrete from the outset, the card below displays our flagship witness from Appendix \ref{app:witness}, resolving a fifteen-year-open public challenge in milliseconds of verification.

\reportwitnesscard{FLAGSHIP SELF-CERTIFYING WITNESS}{Keccak[r=640, c=160, nr=6] $\cdot$ Official keccak.team OK!}{A fifteen-year-open public challenge resolved by two distinct 638-bit single-block messages (Hamming distance 236 bits) that produce the exact same complete 160-bit fingerprint under six rounds of \texttt{Keccak-f[800]}:}{\renewcommand{\arraystretch}{1.08}\begin{tabular}{@{}l@{\hspace{4mm}}l@{}}
{\sffamily\fontsize{7.5}{9.2}\selectfont\bfseries\color{ReportPrimary}Message 1 (638\,b)} & {\scriptsize\ttfamily\color{ReportGray}f78dcbcd8dde8edaa1c49505d530f4f4825a121dafb18995127e0b7e897abacffa7feae79f602e31}\\[0.5pt]
 & {\scriptsize\ttfamily\color{ReportGray}33bd83391fd316d3da6c2a5e49afd3a2aab030d34696fef2e48aed9cf77c0dfa9efda0e52f579a02}\\[4pt]
{\sffamily\fontsize{7.5}{9.2}\selectfont\bfseries\color{ReportPrimary}Message 2 (638\,b)} & {\scriptsize\ttfamily\color{ReportGray}69e855a82fca2cce40dd741cf562d4a6abd83b9fa1b38797b9fca0fc475674e3de75ceedbfe20eb3}\\[0.5pt]
 & {\scriptsize\ttfamily\color{ReportGray}035cb3d88fd386d37d848db678a9e2a41db087d398162072428c4b9a15a2ef2446ef78f77457c102}\\[5pt]
{\sffamily\fontsize{7.5}{9.2}\selectfont\bfseries\color{ReportPrimary}Shared 160-bit digest} & {\setlength{\fboxsep}{2.2pt}\colorbox{ReportLiveBg}{\scriptsize\ttfamily\bfseries\color{ReportPrimary}\,2fa7a934a40df4f99877d0fb3515c0ac58ce2dc7\,}}\enspace{\sffamily\fontsize{7.0}{8.5}\selectfont\bfseries\color{ReportAccent}OK!}
\end{tabular}}{Full parameters and verification details appear in Appendix~\ref{app:witness}. All thirteen verified practical results (twenty-five self-certifying witnesses across Keccak and standard FIPS 197 \texttt{AES-128}, including a 9.17\,s laptop 7-round \texttt{AES-128-MMO} collision) are summarized in Table~\ref{tab:keccakresults} and Appendix~\ref{app:witness}.}

Our broader vision spans two intertwined horizons. On the trajectory of intelligence, we aim to complete cryptanalysis's leap from automated tools to intelligent discovery by establishing an agentic paradigm that hypothesizes, adversarially audits itself, and co-evolves across theory and silicon toward novel attack techniques and eventually new foundational paradigms. On the transformation of cryptography itself, we seek to bring the Oracle out of theoretical security proofs into reality, shifting security from reactive patching under scarce specialist attention toward a proactive science in which every primitive faces continuous, open, self-certifying scrutiny and next-generation standards co-evolve with the oracle before deployment. Within that shift, human cryptographers rise into legislators of security goals, challengers of hidden assumptions, and distillers of mathematical theory. For decades, Alice and Bob dreamed of the Oracle only on the blackboard of security proofs. With openness and mathematical rigor, this is how we begin to forge it in reality.

\Needspace{10\baselineskip}

\section{\texorpdfstring{Wishful Cryptanalysis: A Methodology for Intelligent Discovery}{Wishful Cryptanalysis: A Methodology for Intelligent Discovery}}\label{sec:methodology}

\begin{figure}[tbp]
\centering
\includegraphics[width=\linewidth]{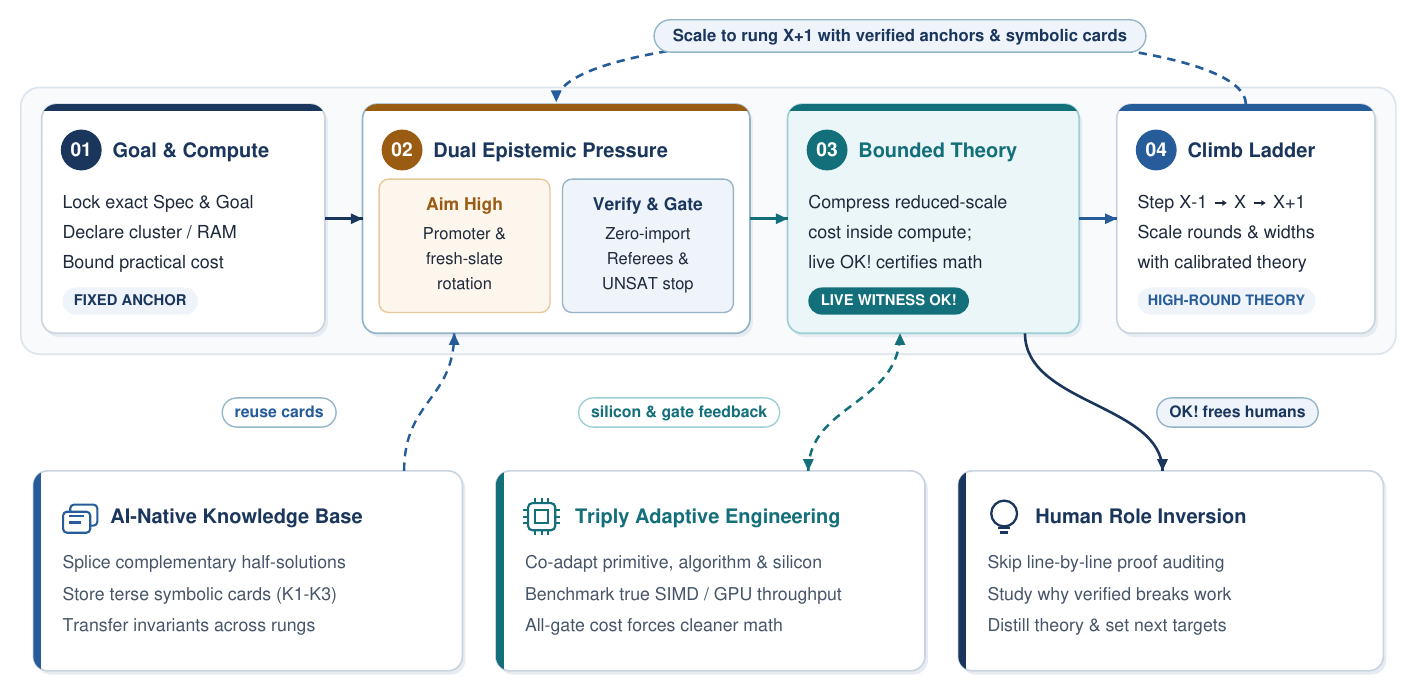}
\caption{The \textbf{Wishful Cryptanalysis} research loop: lock Goal and compute upfront (\texttt{01}), balance constructive optimism against blind zero-import verification and \texttt{UNSAT} stop gates (\texttt{02}), compress reduced-scale complexity into provisioned compute so live \texttt{OK!} witnesses certify LLM math (\texttt{03}), and climb the parameter ladder (\texttt{04}) supported by AI-native knowledge cards, triply adaptive engineering, and human role inversion.}
\label{fig:lifecycle}
\end{figure}

Crossing from automated computation to intelligent discovery requires rethinking how human researchers, foundation models, and formal solvers divide epistemic labor. Standard coding harnesses, whether single agents or flat swarms running generic code-and-test loops, stall on nontrivial cryptanalysis because they treat symbolic derivation, literature precedent, and hardware execution symmetrically. We name the resulting methodology \textbf{wishful cryptanalysis} not for naive dreaming that expects breakthroughs from thin air, but in the classical mathematical spirit of \textit{wishful thinking}. One dares to proceed as if a hidden structural path still exists so as to uncover which implicit assumption can be relaxed, while subjecting every step to compute-bounded execution and blinded, zero-import refutation. Four reinforcing principles operationalize this paradigm, summarized in Figure \ref{fig:lifecycle} and Table \ref{tab:practice}.

\reportcallout{Core Principle of Wishful Cryptanalysis}{Lock the exact target and available compute upfront, then drive models to theoretically compress reduced-scale complexity within that compute window so live witnesses certify their derivations without manual proof-auditing. Pair constructive optimism and existence-first reframing (``aim high'') with zero-import Referees, triply adapt across target primitive, attack algorithm, and hardware, and climb the parameter-and-round ladder to trustworthy asymptotic theory.}

\Needspace{6\baselineskip}

\subsection{\texorpdfstring{Compute-bounded theory, practical verification, and the inversion of human effort}{Compute-bounded theory, practical verification, and the inversion of human effort}}\label{sec:bounded_theory}

Any serious cryptanalytic claim from a language model immediately faces a question of trust: \textit{is the AI's derivation actually right?} Historically, academic cryptanalysis proceeds theory-first and practical-later. Human specialists derive high-round or asymptotic attacks on paper, say at $2^{150}$ complexity, whose credibility rests on human proof-checking while small-scale implementations serve as secondary illustrations. Applied naively to LLMs, that order breaks down. Operating purely on paper at high complexity, models routinely miscount degrees of freedom, overlook boundary conditions, or omit replay overheads, forcing humans to audit dozens of pages of subtle probabilistic algebra by hand.

We invert that order, not by abandoning theory for blind brute force, but by using compute-bounded execution to certify LLM theory. We first task models with theoretical cryptanalysis on reduced-round or reduced-parameter instances under an explicit mandate to compress theoretical complexity inside our provisioned compute budget, and then execute the attack to recover a live witness. Unlike the human-only era, when coding a custom algebraic solver and SIMD engine took weeks, agents write, debug, and run that code themselves. When an independent five-line checker confirms \texttt{OK!}, the witness certifies not only the break, but also that the model's algebraic structure, freedom count, and probability model are genuinely sound. Freed from arithmetic proofreading, human cryptographers study \textit{why} the verified construction works, distill it into lucid prose and compact symbolic invariants, and scale the validated mechanism upward along the parameter ladder (Section \ref{sec:ladder}).

\Needspace{6\baselineskip}

\subsection{\texorpdfstring{Climbing the parameter-and-round ladder}{Climbing the parameter-and-round ladder}}\label{sec:ladder}

Practical verification is not a ceiling, but a launchpad. Following the spirit of structured challenge ladders such as the Keccak Crunchy Contest \citep{crunchy} or lattice and decoding challenges, we set explicit step-by-step targets across rounds, state widths, or algebraic dimensions. Solving rung $X-1$ practically forces the agent team to confront every real specification detail, from domain bits and initial state constraints to key-schedule couplings and memory-hierarchy limits.

Once those mechanics are pinned down by a live witness at rung $X-1$, extrapolating to rung $X$, wider states, or full standard security levels changes character completely. High-complexity derivation is no longer ungrounded speculation. Every degree count, subspace dimension, and transition probability is calibrated against an executable baseline one step below, making high-round and asymptotic theory trustworthy.

\Needspace{6\baselineskip}

\subsection{\texorpdfstring{Disarming dual hallucinations: constructive optimism vs. zero-import Referees}{Disarming dual hallucinations: constructive optimism vs. zero-import Referees}}\label{sec:hallucinations}

The two epistemic hallucinations identified in Section \ref{sec:opportunities} pull in opposite directions, one causing agents to quit before starting and the other inflating unverified claims. Disarming them therefore requires opposite, strictly separated interventions.

\begin{itemize}

\item \textbf{Countering authority hallucination via constructive optimism and fresh-slate rotation.} When a model reads an influential paper or an earlier session log stating that an avenue ``hits a structural wall,'' it mistakes an obstacle of \textit{one restricted ansatz} (such as requiring symmetric input messages or controlling at most three linear layers) for an unconditional impossibility of the primitive itself. Whereas humans rarely have time to re-test ten discarded strategies, AI has the capacity to explore territory prior attempts left behind, provided it does not surrender first. To give exploring agents the impetus to push past learned deference to negative precedent, we pair \textit{constructive optimism} with \textit{existence-first reframing}, treating solvability as an active working conviction (\textit{``a practical path still exists at this round; an implicit assumption in the stalled route can be relaxed, so dig deeper''}). We back that posture with fresh-slate specialist rotation, spawning a new theorist stripped of prior failed branch histories whenever active threads share an entrenched coordinate blind spot. We call this spirit aiming high to harvest real breakthroughs (\textit{qi\'u sh\`ang d\'e zh\={o}ng}). Even when an ambitious target ultimately resists a practical solve, venturing past superficial ``no-go'' lore uncovers overlooked degrees of freedom and sharper sub-generic bounds.

\item \textbf{Countering reasoning hallucination via zero-import Referees and micro-experiments.} Bold exploration must never relax verification or burn compute blindly on dead ends. Every theoretical derivation, complexity claim, and solver script faces zero-peer-import adversarial Referees, strict specification-legality and all-Boolean-gate cost rulers, and discriminating micro-experiments. Whenever specialists disagree on a subspace dimension, elimination deficit, or tail floor, the workflow replaces verbal debate with decisive computation such as exact symbolic rank probes or exhaustive state-kernel enumerations in native C++. That calculation provides a hard stopping gate (\texttt{UNSAT} or a certified lower bound) once a genuine mathematical wall is reached.

\end{itemize}

\Needspace{6\baselineskip}

\subsection{\texorpdfstring{Continuous 24/7 evolution and triply adaptive engineering}{Continuous 24/7 evolution and triply adaptive engineering}}\label{sec:adaptive}

Unlike human investigators who must context-switch and rest, an agent team operates twenty-four hours a day so long as tokens and compute are provisioned. Once theoretical structure matures, agents move beyond black-box solvers to \textbf{triply adaptive engineering}, simultaneously adapting to the target cryptographic primitive, the attack algorithm, and the provisioned silicon. Compiling primitive-specific S-box flats or ring arithmetic, reshaping data structures around the analytical paradigm, and tuning SIMD registers, cache layouts, and GPU warps to the underlying hardware turns token budgets and compute hours into compounding speedups.

\begin{table}[tbp]
\caption{Core bottlenecks in autonomous cryptanalysis and how wishful cryptanalysis resolves them.}
\label{tab:practice}
{\small
\setlength{\aboverulesep}{0pt}\setlength{\belowrulesep}{0pt}
\setlength{\extrarowheight}{2.6pt}
\renewcommand{\arraystretch}{1.22}
\setlength{\tabcolsep}{6.0pt}
\begin{tabularx}{\linewidth}{@{}>{\raggedright\arraybackslash}p{.185\linewidth}>{\raggedright\arraybackslash}p{.295\linewidth}>{\raggedright\arraybackslash}X@{}}
\arrayrulecolor{ReportPrimary}\toprule[1.1pt]
\rowcolor{ReportHeaderBg}{\leavevmode\sffamily\bfseries\color{ReportPrimary}Bottleneck} & {\leavevmode\sffamily\bfseries\color{ReportPrimary}Failure in naive harnesses} & {\leavevmode\sffamily\bfseries\color{ReportPrimary}Wishful cryptanalysis resolution} \\
\arrayrulecolor{ReportAccent}\midrule[0.65pt]
{\leavevmode\bfseries\color{ReportPrimary}Authority hallucination} & Models surrender when prior literature deems a restricted ansatz blocked. & Apply constructive optimism (\textit{``relax an assumption''}), \textbf{fresh-slate rotation}, and a \textbf{Promoter}. \\
\arrayrulecolor{ReportRowRule}\specialrule{0.35pt}{0pt}{0pt}
\rowcolor{ReportZebra}{\leavevmode\bfseries\color{ReportPrimary}Reasoning hallucination} & Models miscount degrees of freedom, omit overheads, or drift out of scope. & Enforce \textbf{zero-import Referees}, \textbf{all-gate \& scope audits}, and decisive \texttt{UNSAT} tests. \\
\arrayrulecolor{ReportRowRule}\specialrule{0.35pt}{0pt}{0pt}
{\leavevmode\bfseries\color{ReportPrimary}Proof-checking burden} & Theory-first paper proofs force humans to audit subtle algebraic bugs by hand. & Compress \textbf{reduced-scale theory into compute} first; live \texttt{OK!} witnesses certify math before scaling. \\
\arrayrulecolor{ReportRowRule}\specialrule{0.35pt}{0pt}{0pt}
\rowcolor{ReportZebra}{\leavevmode\bfseries\color{ReportPrimary}Paper-to-silicon gap} & Generic solvers lag by orders of magnitude; nominal steps mislead real cost. & Dispatch \textbf{Engineers ($1\dots N$)} to \textbf{triply adapt} across primitive, algorithm, and silicon. \\
\arrayrulecolor{ReportRowRule}\specialrule{0.35pt}{0pt}{0pt}
{\leavevmode\bfseries\color{ReportPrimary}Fragmented insights} & Parallel threads discover complementary front and tail pieces in isolation. & Pair a topic \texttt{@}-board with \textbf{Knowledge Stewards} splicing half-wins into symbolic cards. \\
\arrayrulecolor{ReportRowRule}\specialrule{0.35pt}{0pt}{0pt}
\rowcolor{ReportZebra}{\leavevmode\bfseries\color{ReportPrimary}Closed platforms \& over-refusal} & General-purpose guardrails on closed platforms often interrupt boundary work. & Use \textbf{open-weight models} and models without over-refusal via an \textbf{open-source framework} and local solvers. \\
\arrayrulecolor{ReportPrimary}\bottomrule[1.0pt]
\end{tabularx}}
\end{table}

\Needspace{10\baselineskip}

\section{\texorpdfstring{Architecture of the Open-Source Cryptanalysis Agent Framework}{Architecture of the Open-Source Cryptanalysis Agent Framework}}\label{sec:architecture}

\begin{figure}[tbp]
\centering
\includegraphics[width=\linewidth]{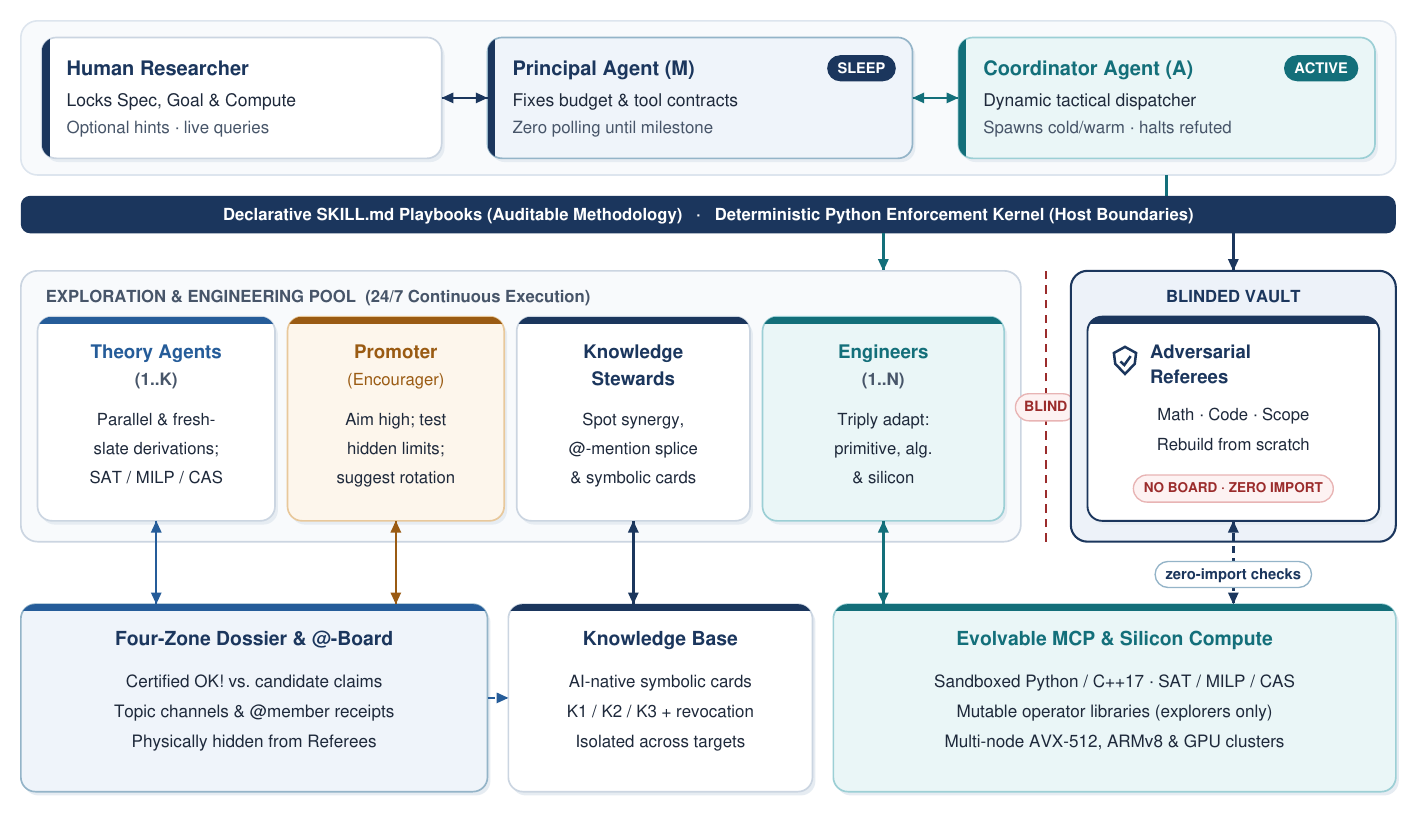}
\caption{Architecture of the open-source \textbf{Wishful Cryptanalysis Agent Framework} (\texttt{agent-\allowbreak{}framework}): declarative \texttt{SKILL.md} playbooks separated from a deterministic Python enforcement kernel, with a sleeping Principal (\texttt{M}), an active Coordinator (\texttt{A}), an exploration/engineering pool, and an explicit \textbf{Blind Isolation Wall} preventing Adversarial Referees from accessing the board or importing explorer code.}
\label{fig:architecture}
\end{figure}

To turn wishful cryptanalysis from personal craft into a shared community capability, we formalized its institutional principles into an open-source multi-agent reference platform (\texttt{agent-\allowbreak{}framework}), while emphasizing that the methodology itself is harness-agnostic and can be instantiated inside \textit{any} agent runtime. Why can frontier cryptanalysis not simply adopt the flat collaboration model of software-engineering swarms? In software engineering, shared unit tests align code merges, so full peer visibility suffices. In scientific discovery, and cryptanalysis in particular, shared discussion easily breeds premature consensus or shared blind spots. Frontier discovery therefore demands \textbf{asymmetric epistemic governance}, an institutional separation of powers that grants explorers bold, assumption-breaking priors while binding critics behind a blind, zero-import verification wall. Our architecture enforces this governance by separating declarative \texttt{SKILL.md} role playbooks (human-auditable cognitive methodology) from a deterministic Python enforcement kernel (Figure \ref{fig:architecture} and Table \ref{tab:roles}).

\begin{table}[tbp]
\caption{Division of responsibilities in the open-source Wishful Cryptanalysis multi-agent architecture.}
\label{tab:roles}
{\small
\setlength{\aboverulesep}{0pt}\setlength{\belowrulesep}{0pt}
\setlength{\extrarowheight}{2.6pt}
\renewcommand{\arraystretch}{1.22}
\setlength{\tabcolsep}{6.0pt}
\begin{tabularx}{\linewidth}{@{}>{\raggedright\arraybackslash}p{.240\linewidth}>{\raggedright\arraybackslash}p{.272\linewidth}>{\raggedright\arraybackslash}X@{}}
\arrayrulecolor{ReportPrimary}\toprule[1.1pt]
\rowcolor{ReportHeaderBg}{\leavevmode\sffamily\bfseries\color{ReportPrimary}Role} & {\leavevmode\sffamily\bfseries\color{ReportPrimary}Activation mode} & {\leavevmode\sffamily\bfseries\color{ReportPrimary}Core responsibility \& enforced boundary} \\
\arrayrulecolor{ReportAccent}\midrule[0.65pt]
{\leavevmode\bfseries\color{ReportPrimary}Principal Agent (\texttt{M})} & Start; \textbf{sleeps} until milestone & Locks Spec, Goal, \textbf{compute envelope}, and tools; zero polling tokens while asleep. \\
\arrayrulecolor{ReportRowRule}\specialrule{0.35pt}{0pt}{0pt}
\rowcolor{ReportZebra}{\leavevmode\bfseries\color{ReportPrimary}Coordinator (\texttt{A})} & Continuous tactical dispatch & Spawns parallel theorists (cold fresh-slate \& warm branches) and halts refuted routes. \\
\arrayrulecolor{ReportRowRule}\specialrule{0.35pt}{0pt}{0pt}
{\leavevmode\bfseries\color{ReportPrimary}Theory Agents ($1\dots K$)} & Independent start, then board & Survey public literature autonomously, derive attacks, and invoke SAT/MILP/CAS. \\
\arrayrulecolor{ReportRowRule}\specialrule{0.35pt}{0pt}{0pt}
\rowcolor{ReportZebra}{\leavevmode\bfseries\color{ReportPrimary}Promoter Agent} & On route stall or stop check & Challenges hidden limits and triggers fresh-slate rotation; cannot approve claims or alter Spec. \\
\arrayrulecolor{ReportRowRule}\specialrule{0.35pt}{0pt}{0pt}
{\leavevmode\bfseries\color{ReportPrimary}Adversarial Referees} & Blinded vault; zero imports & Math, Code, and Scope critics rebuild checks from scratch for legality and all-gate cost. \\
\arrayrulecolor{ReportRowRule}\specialrule{0.35pt}{0pt}{0pt}
\rowcolor{ReportZebra}{\leavevmode\bfseries\color{ReportPrimary}Engineers ($1\dots N$)} & Once theory passes audit & \textbf{Triply adapt} code across target primitive, attack algorithm, and silicon clusters. \\
\arrayrulecolor{ReportRowRule}\specialrule{0.35pt}{0pt}{0pt}
{\leavevmode\bfseries\color{ReportPrimary}Knowledge Stewards} & On blockers or partial wins & Splice complementary half-solutions on the board into AI-native cards (\texttt{K1}--\texttt{K3}). \\
\arrayrulecolor{ReportPrimary}\bottomrule[1.0pt]
\end{tabularx}}
\end{table}

\Needspace{6\baselineskip}

\subsection{\texorpdfstring{Anchoring target goal and compute resources upfront}{Anchoring target goal and compute resources upfront}}\label{sec:anchoring}

Launching an investigation locks down two non-negotiable anchors, the exact Goal and the provisioned compute envelope. First, the human researcher supplies the primitive specification (mathematical definition or reference C/Python code) and fixes the target instance, parameter or round regime, access model, and victory condition. This gives agents an unambiguous bullseye so exploration never drifts into weakened variants. Second, declaring available compute and memory resources (from laptop silicon and RAM limits to multi-server AVX-512 and GPU clusters) calibrates \textit{what theoretical complexity threshold counts as practically verifiable}. Theorists learn from the outset whether reduced-scale derivations must be compressed below $2^{32}$, $2^{40}$, or $2^{52}$ to achieve live witness certification.

Crucially, human attack directions remain optional. If a cryptanalyst supplies a hunch or seed papers, the system expands from there. Even with \textit{zero} hints, agents autonomously search public cryptographic literature, inspect prior attacks, audit reference benchmarks, and synthesize initial avenues on their own.

\Needspace{6\baselineskip}

\subsection{\texorpdfstring{Sleeping Principal and active Coordinator}{Sleeping Principal and active Coordinator}}\label{sec:control_plane}

Multi-agent systems fail at two extremes. Flat group chats drift without convergence, while a single central controller monitoring every tool call exhausts its context window and token budget. We resolve this tension through a two-tier control hierarchy.

\begin{itemize}

\item \textbf{Principal Agent (M), strategic anchor in sleep mode.} At session start, the Principal locks down the exact Goal, verifies test vectors, binds the declared compute envelope, practical complexity target, and base tool contracts, and delegates the research agenda to the Coordinator. The Principal then enters sleep mode, consuming zero polling tokens. It wakes \textit{only} on two triggers, namely when a human researcher checks in conversationally (e.g., \textit{``Where do we stand right now?''} or \textit{``Add the GPU cluster and push further''}), or when the Coordinator reports a verified milestone breakthrough or requests a session-level stopping decision.

\item \textbf{Coordinator Agent (A), dynamic tactical commander.} Operating continuously on the frontier, the Coordinator spawns multiple Theory Subagents with distinct initial questions so independent derivations form before group discussion begins. Watching emerging evidence and discussion, it dynamically spawns additional theoretical directions, pairing cold fresh-slate specialists (initialized solely from the raw specification whenever existing threads share a blind spot) with warm-forked branches (inheriting verified computational receipts to explore parameter splits), and terminates subagents whose specific assumptions have been conclusively refuted.

\end{itemize}

\Needspace{6\baselineskip}

\subsection{\texorpdfstring{Institutionalizing optimism and skepticism: Promoter and Referees}{Institutionalizing optimism and skepticism: Promoter and Referees}}\label{sec:epistemic_roles}

The two opposite pressures of wishful cryptanalysis are embodied as first-class, permission-separated roles enforced by the host runtime.

\begin{itemize}

\item \textbf{Promoter Agent (the Encourager).} When theoretical subagents report barriers or the Principal reaches a stopping checkpoint, the Promoter inspects the failed routes. Rather than emitting generic slogans, it challenges implicit restrictions (\textit{``Did the target specification require symmetric inputs, or only our initial ansatz?''}), injects constructive confidence to revisit discarded strategies, recommends fresh-slate specialist rotation when prior branches share an unexamined parameterization, and proposes cheap discriminating experiments to test whether a ``blocked'' avenue still harbors viable subcases. It cannot approve claims or alter the target. Its sole role is to prevent authority hallucination from closing an open door.

\item \textbf{Independent Adversarial Referees (Math, Code, and Scope Critics).} Every mathematical claim, solver encoding, complexity count, and attack program faces dedicated Math, Code, and Scope Referees. Each referee runs in a fresh, physically blinded context containing only the exact target specification, frozen claim, and primary artifacts, never the team's wishful framing, human nudges, or discussion board. Crucially, the runtime enforces zero-peer-import verification, blocking referees from importing session operators authored by explorers and forcing them to reconstruct the target primitive, boundary rules, and rank checks from scratch. They audit parameter legality (distinguishing nominal specification windows from extended diagnostic regimes) and all-Boolean-gate complexity (charging every preprocessing, linear-algebra, table-unpack, and tail gate rather than isolated remaining-round calls). Blocking objections register structured obligations that must be resolved before any claim advances.

\end{itemize}

\Needspace{6\baselineskip}

\subsection{\texorpdfstring{Triply adaptive Engineer Agents and self-evolving MCP infrastructure}{Triply adaptive Engineer Agents and self-evolving MCP infrastructure}}\label{sec:mcp_engineers}

Before exploration begins, the host exposes a unified computational substrate via the Model Context Protocol (MCP) \citep{mcp}. It bundles sandboxed Python and native C++17, hash-pinned witness oracles, proof-checked SAT, exact-integer MILP, and symbolic algebra, paired with extensible tool-skill and cluster adapters for external solvers and multi-node CPU/GPU servers. Within host permission boundaries, cryptographic operator code remains open to adaptive evolution across sessions.

Once theoretical exploration and referee review settle on a sound attack structure, the Coordinator dispatches multiple named Engineer Agents ($1\dots N$) for three-way adaptation. They adapt to the target primitive (compiling custom S-box flats, key-bridge tables, or zero-XOR projections rather than generic encodings), to the attack algorithm (tailoring structures to internal differentials, Gray-code polynomial evaluation, streaming meet-in-the-middle, or multi-slot distinguished-point walks), and to the provisioned silicon (matching coset dimensions, AVX-512 or ARMv8 Crypto intrinsics, and CUDA warps to cache and cluster RAM). Real hardware throughput recalibrates theoretical parameters. Equally important, strict all-Boolean-gate accounting actively forces cleaner mathematics, replacing runtime Gaussian elimination and lookup tables with pure-\texttt{\{XOR\}} algebraic recurrences and symplectic normal forms.

\Needspace{6\baselineskip}

\subsection{\texorpdfstring{Public \texttt{@}-mention board and Knowledge Stewards for AI-native synthesis}{Public @-mention board and Knowledge Stewards for AI-native synthesis}}\label{sec:knowledge_stewards}

Hard cryptanalytic breakthroughs rarely emerge inside a single thread. Frequently, Subagent 1 discovers how to linearize front rounds while Subagent 2 independently characterizes a low-dimensional output subspace at the tail. Neither constitutes a full attack alone, yet their combination breaks the target. Enabling that fusion requires two mechanisms.

\begin{enumerate}

\item \textbf{Persistent Topic Board with \texttt{@}-Mentions and Four-Zone Auditable Dossiers.} Once initial independent hypotheses are posted, active theorists and engineers collaborate across topic-scoped channels and directed \texttt{@member} handoffs, attaching exact equations, basis files, or counterexamples with host-tracked delivery receipts. Simultaneously, the host structures every session into an auditable four-zone dossier that strictly separates (i) the frozen target specification, (ii) the shared board distinguishing host-certified \texttt{OK!} findings from unverified candidate claims, (iii) each specialist's declarative \texttt{SKILL.md} playbook and local record, and (iv) physically isolated Referee and primary-evidence vaults.

\item \textbf{Knowledge Steward Agents and AI-Native Symbolic Cards.} Invoked when routes hit blockers or produce partial components, Knowledge Stewards scan the collective record to spot complementary half-solutions, \texttt{@}-mention the relevant specialists to fuse them, and curate verified insights into a Project Knowledge Base (stratified into mathematical lemmas \texttt{K1}, structural techniques \texttt{K2}, and instance results \texttt{K3}, with dependency-tracked revocation and strict isolation across target families). Internal cards are stored as concise symbolic rules that peer agents parse and transfer instantaneously across rungs:

\end{enumerate}

\reportkcard{KNOWLEDGE CARD K2-SPONGE-SYM-04}{Scope: Permutation Sponges  $\cdot$  Status: Verified}{\textbf{1. Symmetry Equivariance:} Under period-$I$ lane folding $\delta_I$, the unkeyed round core commutes ($[\delta_I, \chi \circ \pi \circ \rho \circ \theta] = 0$); asymmetry enters $S_I = \ker(\delta_I)$ solely through round-constant XOR $\delta_I(\text{RC})$.\par\vspace{2.5pt}\textbf{2. Quadratic-to-Affine Connector Rule:} Restricting active front-round S-boxes to 2D affine flats keeps the next round degree-1 in free message variables, eliminating quadratic blowup across two full rounds.\par\vspace{2.5pt}\textbf{3. Boundary Difference / Value Decoupling:} On boundary lanes mixing message and fixed padding bits, paired bits remain free in Stage-1 differences while freezing Stage-2 symmetric values, reused across 6-round \texttt{Keccak[640, 160]} and equal-length \texttt{b=1600} sponge solves.}

\Needspace{10\baselineskip}

\section{\texorpdfstring{Frontier Campaigns: Practical Results on Keccak and FIPS 197 AES-128}{Frontier Campaigns: Practical Results on Keccak and FIPS 197 AES-128}}\label{sec:campaigns}

To test whether wishful cryptanalysis generalizes across fundamentally different design philosophies, we conducted sustained campaigns on both pillars of modern symmetric-key cryptography. We chose cryptographic hash functions, the foundational primitives that compress arbitrary messages into unforgeable digital fingerprints, as our primary arena. Finding a \textit{collision} means discovering two distinct messages with the same fingerprint, while finding a \textit{preimage} means inverting a given fingerprint back to an input. Our campaigns spanned unkeyed sponge permutations in the Keccak Crunchy Crypto Collision and Pre-image Contest \citep{crunchy,sha3}, alongside key-scheduled wide-trail block-cipher hashing on standard FIPS 197 \texttt{AES-128} \citep{aes} in both Matyas--Meyer--Oseas (\texttt{MMO}, $H_i = \mathrm{AES\text{-}128}_{H_{i-1}}^{(n_r)}(M_i) \oplus M_i$) and Davies--Meyer (\texttt{DM}, $H_i = \mathrm{AES\text{-}128}_{M_i}^{(n_r)}(H_{i-1}) \oplus H_{i-1}$) modes. Formally, \texttt{Keccak-f[b]} iterates $\iota_r \circ \chi \circ \pi \circ \rho \circ \theta$ over a $5\times 5\times w$ state ($b = 25w \in \{200, 400, 800, 1600\}$), absorbing $r$-bit blocks at zero initial $c$-bit capacity ($r+c=b$) under \texttt{pad10*1}. Unmodified $n_r$-round \texttt{AES-128} applies \texttt{SubBytes}, \texttt{ShiftRows}, \texttt{MixColumns} (omitted in round $n_r$), and nonlinear \texttt{SubWord} key expansion over $4\times 4$ byte states.

Across diverse frontier open-weight and proprietary foundation models,\footnote{Campaigns spanned multiple frontier open-weight and proprietary checkpoints, including DeepSeek v4.1 Flash, GLM 5.3, Kimi 3.1 Pro, ChatGPT 6 Astra \& 6.1 Sol, Claude Opus 4.8 \& 5.5, and Gemini 4 Argon, Carbon \& Neon.} agents carried out all mathematical derivations, solver encodings, and hardware optimizations autonomously, incurring under \$1,000 USD in cumulative LLM inference cost across all thirteen results (separate from downstream execution compute; Section \ref{sec:rigor_flywheel}). Human involvement remained restricted to locking exact target specifications, provisioning compute, and offering occasional high-level reframing when a route stalled (as in Section \ref{sec:keccak_b800}), demonstrating both the model-agnostic reach and low financial barrier of this paradigm. Consistent with Section \ref{sec:open_oracle}, general-purpose guardrails on closed commercial platforms frequently interrupted sustained boundary research mid-derivation, making open-weight models and research-friendly endpoints markedly more reliable in practice.

Following the compute-bounded certification principle of Section \ref{sec:bounded_theory}, this monograph restricts its reported claims strictly to thirteen verified practical results backed by twenty-five self-certifying witnesses in Appendix \ref{app:witness} (Figure \ref{fig:keccak} and Table \ref{tab:keccakresults}). These results are anchored by an official fifteen-year leaderboard record and headline practical firsts across both sponge and \texttt{AES-128} tracks, with extended-window preimages explicitly demarcated as \texttt{Live (Ext.)}. All accompanying theoretical complexity bounds, high-round derivations, and machine-checked obstruction/UNSAT certificates (spanning \texttt{SHA3-384}, \texttt{Keccak[240, 160]}, nominal 3-round \texttt{b=200} and 5-round \texttt{b=1600} preimages, and next-round \texttt{AES-128} walls) are archived in full alongside runnable kernels on our public repository and companion technical papers.

\begin{figure}[tbp]
\centering
\includegraphics[width=\linewidth]{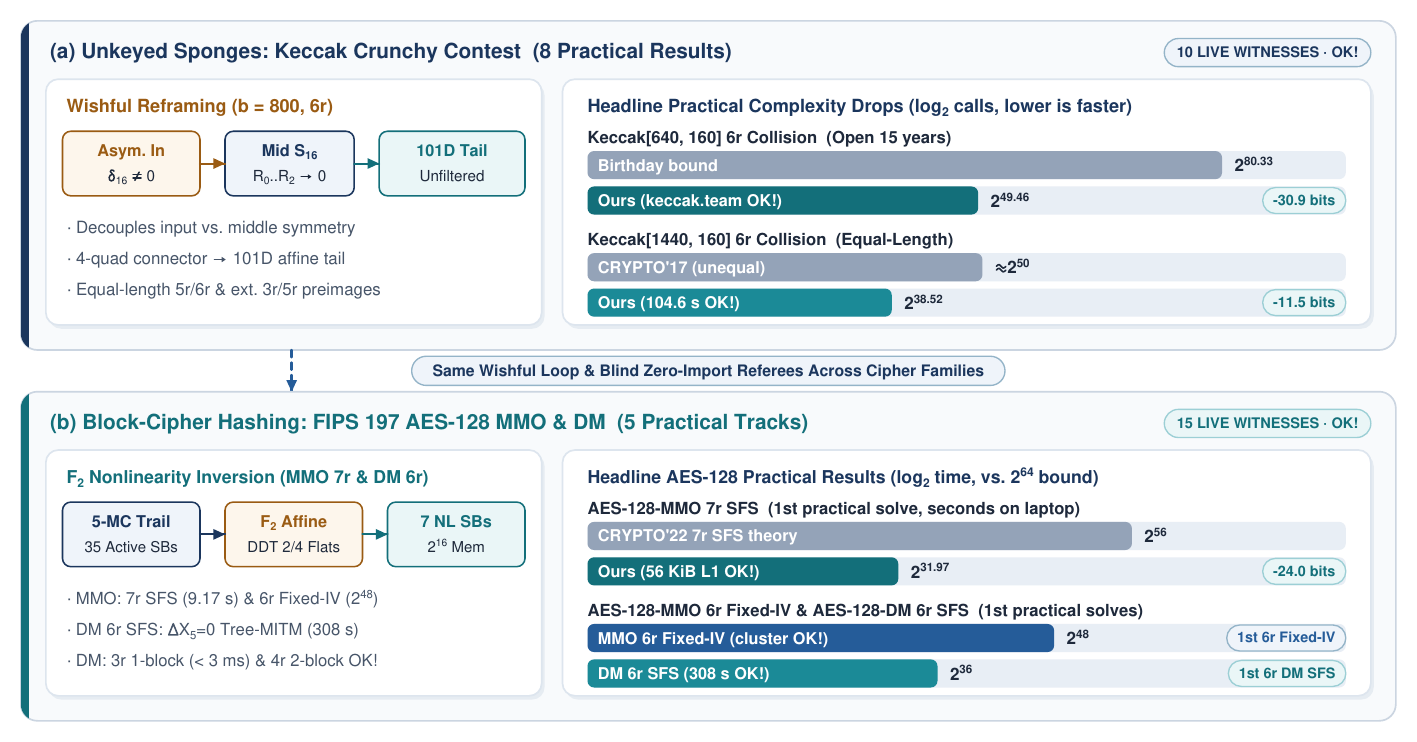}
\caption{Two practical frontier campaigns of \textbf{Wishful Cryptanalysis} (high-level mechanisms and $\log_2$ complexity drops): \textbf{(a) Unkeyed Sponges (\texttt{Keccak})}, breaking 15-year-open \texttt{Keccak[640, 160, 6]} ($2^{80.33}\to 2^{49.46}$, \texttt{keccak.team} \texttt{OK!}) and reducing \texttt{Keccak[1440, 160, 6]} by over three orders of magnitude to $2^{38.52}$ with equal-length live witnesses; and \textbf{(b) Standard FIPS 197 \texttt{AES-128} (\texttt{MMO} \& \texttt{DM})}, sweeping all five collision tracks with the first practical 7-round \texttt{MMO} SFS ($2^{31.97}$ vs. CRYPTO 2022 $2^{56}$ SFS bound), 6-round \texttt{MMO} Fixed-IV ($2^{48}$), and 6-round \texttt{DM} SFS ($308\text{ s}$, $2^{36}$) collisions.}
\label{fig:keccak}
\end{figure}

\begin{table}[tbp]
\caption{Summary of thirteen verified practical cryptanalytic results (twenty-five self-certifying witnesses) across Keccak sponges and all five standard FIPS 197 AES-128 hashing tracks (MMO and DM).}
\label{tab:keccakresults}
{\footnotesize
\setlength{\aboverulesep}{0pt}\setlength{\belowrulesep}{0pt}
\setlength{\extrarowheight}{1.2pt}
\renewcommand{\arraystretch}{1.08}
\setlength{\tabcolsep}{4.2pt}
\begin{tabularx}{\linewidth}{@{}>{\centering\arraybackslash}p{.032\linewidth}>{\raggedright\arraybackslash}p{.232\linewidth}>{\raggedright\arraybackslash}p{.220\linewidth}>{\raggedright\arraybackslash}X>{\centering\arraybackslash}p{.106\linewidth}@{}}
\arrayrulecolor{ReportPrimary}\toprule[1.1pt]
\rowcolor{ReportHeaderBg}{\leavevmode\sffamily\bfseries\color{ReportPrimary}\#} & {\leavevmode\sffamily\bfseries\color{ReportPrimary}Target} & {\leavevmode\sffamily\bfseries\color{ReportPrimary}Prior SOTA} & {\leavevmode\sffamily\bfseries\color{ReportPrimary}Headline live result \& high-level mechanism} & {\leavevmode\sffamily\bfseries\color{ReportPrimary}Status} \\
\arrayrulecolor{ReportAccent}\midrule[0.65pt]
\rowcolor{ReportGroupBg}\multicolumn{5}{l}{\leavevmode\sffamily\bfseries\fontsize{8.0}{10.5}\selectfont\color{ReportPrimary}Part A --- Unkeyed Sponge Permutations: Keccak Crunchy Contest (8 Practical Results $\cdot$ 10 Witnesses)} \\
\arrayrulecolor{ReportRule}\specialrule{0.45pt}{0pt}{0pt}
1 & \texttt{Keccak[640, 160]} 6r Coll & Open 15 yrs ($2^{80.33}$) & \textbf{First practical 6r 160b collision} ($2^{49.46}$ calls; \texttt{keccak.team} leaderboard) via asymmetric-to-$S_{16}$ connector \& 101D tail. & \tabbadge{ReportLiveBg}{ReportAccent}{Live OK!} \\
\arrayrulecolor{ReportRowRule}\specialrule{0.35pt}{0pt}{0pt}
\rowcolor{ReportZebra}2 & \texttt{Keccak[640, 160]} 5r Coll & 5r unequal \citep{qiao17} & \textbf{First single-block equal-length 5r collision} ($b=800$, $638=638\text{ b}$); 2-block full collision in $9.44\text{ ms}$. & \tabbadge{ReportLiveBg}{ReportAccent}{Live OK!} \\
\arrayrulecolor{ReportRowRule}\specialrule{0.35pt}{0pt}{0pt}
3 & \texttt{Keccak[1440, 160]} 6r Coll & 6r unequal $\approx 2^{50}$ \citep{song17} & \textbf{$>10^3\times$ lower complexity ($2^{38.52}$)} covering all 19 legal windows; \textbf{first equal-length 6r live witnesses} at \texttt{start=0} ($104.6\text{ s}$) \& \texttt{18} ($146.6\text{ s}$). & \tabbadge{ReportLiveBg}{ReportAccent}{Live OK!} \\
\arrayrulecolor{ReportRowRule}\specialrule{0.35pt}{0pt}{0pt}
\rowcolor{ReportZebra}4 & \texttt{Keccak[1440, 160]} 5r Coll & 5r unequal \citep{qiao17} & \textbf{First equal-length 5r collision ($0.899\text{ ms}$)} \& 2-block full 1600-bit state collision ($0.187\text{ ms}$). & \tabbadge{ReportLiveBg}{ReportAccent}{Live OK!} \\
\arrayrulecolor{ReportRowRule}\specialrule{0.35pt}{0pt}{0pt}
5 & \texttt{Keccak[40, 160]} 2r Coll & Relied on \texttt{RC[3]=0} & \textbf{First all-nonzero-RC full-state 2r collision} (\texttt{start=9}, $b=200$) solved live in $10.37\text{ min}$. & \tabbadge{ReportLiveBg}{ReportAccent}{Live OK!} \\
\arrayrulecolor{ReportRowRule}\specialrule{0.35pt}{0pt}{0pt}
\rowcolor{ReportZebra}6 & \texttt{Keccak[40, 160]} 2r Pre & Leaderboard $517\text{ b}$ & \textbf{477-bit preimage (12 blocks) for the official 2r target} ($40\text{ bits}$ shorter than the leaderboard entry, $< 16\text{ MiB}$ RAM). & \tabbadge{ReportLiveBg}{ReportAccent}{Live OK!} \\
\arrayrulecolor{ReportRowRule}\specialrule{0.35pt}{0pt}{0pt}
7 & \texttt{Keccak[40, 160]} 3r Pre & Open ($2^{80.00}$) & \textbf{Extended LFSR window (\texttt{start=93}) live 80b preimage} ($398\text{ b}$, solved in $2\text{ h }37\text{ min}$; separated from nominal \texttt{0..15}). & \tabbadge{ReportAmberBg}{ReportAmber}{Live (Ext.)} \\
\arrayrulecolor{ReportRowRule}\specialrule{0.35pt}{0pt}{0pt}
\rowcolor{ReportZebra}8 & \texttt{Keccak[1440, 160]} 5r Pre & Open ($2^{80.00}$) & \textbf{Extended LFSR window (\texttt{start=73}) live 80b preimage} ($5{,}758\text{ b}$ / 4-blk, homogeneous $H_6$ Yule MitM, \textbf{$5\text{ h }35\text{ min}$}; separated from \texttt{0..19}). & \tabbadge{ReportAmberBg}{ReportAmber}{Live (Ext.)} \\
\arrayrulecolor{ReportRule}\specialrule{0.55pt}{0pt}{0pt}
\rowcolor{ReportGroupBg}\multicolumn{5}{l}{\leavevmode\sffamily\bfseries\fontsize{8.0}{10.5}\selectfont\color{ReportPrimary}Part B --- Block-Cipher Hashing: Standard FIPS 197 AES-128 MMO \& DM (5 Practical Tracks $\cdot$ 15 Witnesses)} \\
\arrayrulecolor{ReportRule}\specialrule{0.45pt}{0pt}{0pt}
9 & \texttt{AES-128-MMO} 7r SFS & 6r prac.; 7r $2^{56}$ \citep{dong22} & \textbf{First practical 7r SFS collision}: \textbf{$(2^{31.97}, 2^{16})$} via 7-NL-SB $\mathbb{F}_2$ 5-\texttt{MC} inbound; \textbf{live on laptop} ($2^{33.20}$ pairs, $9.17\text{ s}$ hit; +6r in $1.55\text{ s}$). & \tabbadge{ReportLiveBg}{ReportAccent}{Live OK!} \\
\arrayrulecolor{ReportRowRule}\specialrule{0.35pt}{0pt}{0pt}
\rowcolor{ReportZebra}10 & \texttt{AES-128-MMO} 6r Fixed-IV & 7r theory \citep{cglsz25}; 6r prac. open & \textbf{First practical 6r Fixed-IV collision} (\texttt{AES-\allowbreak{}CRUNCHY-\allowbreak{}v1}, $2^{48}$ $\Delta Z_2$-factored Super-Sbox \citep{gp10}); \textbf{5r live at $2^{32}$} ($0.052\text{ s}$ / $23.52\text{ s}$, +4r). & \tabbadge{ReportLiveBg}{ReportAccent}{Live OK!} \\
\arrayrulecolor{ReportRowRule}\specialrule{0.35pt}{0pt}{0pt}
11 & \texttt{AES-128-DM} 6r SFS & 5r prac. \citep{taiyama24,qin25}; 6r open & \textbf{First practical 6r DM SFS collision}: $\Delta X_5=0$ + partial-$Z_1$ $\mathbb{F}_2$ Tree-MITM + \texttt{0xc5} ($2^{36}$, $32\text{--}64\text{ KiB}$); \textbf{live in $308\text{ s}$} (+4r/5r). & \tabbadge{ReportLiveBg}{ReportAccent}{Live OK!} \\
\arrayrulecolor{ReportRowRule}\specialrule{0.35pt}{0pt}{0pt}
\rowcolor{ReportZebra}12 & \texttt{AES-128-DM} 3r 1-Blk IV & 3r prac. \citep{qin25}; 4r theory \citep{cglsz25} & \textbf{Dual-season autonomous live solves} ($\sim 2^{21.5}$ cascade rebound): \textbf{$0.64\text{ ms}$} (\texttt{ZERO-IV}) \& \textbf{$2.65\text{ ms}$} (\texttt{AES-\allowbreak{}CRUNCHY-\allowbreak{}v1}). & \tabbadge{ReportLiveBg}{ReportAccent}{Live OK!} \\
\arrayrulecolor{ReportRowRule}\specialrule{0.35pt}{0pt}{0pt}
13 & \texttt{AES-128-DM} 4r 2-Blk IV & 3r prac. \citep{taiyama24}; 4r $H_1{=}H_1^\prime$ \citep{tang26} & \textbf{Dual-season autonomous 2-block solves ($H_1\neq H_1^\prime$)} via 64b DP birthday ($2^{36.5}$, $96\text{ MiB}$) + MDS inversion: \textbf{$157.0\text{ s}$} / \textbf{$589.8\text{ s}$}. & \tabbadge{ReportLiveBg}{ReportAccent}{Live OK!} \\
\arrayrulecolor{ReportPrimary}\bottomrule[1.0pt]
\end{tabularx}}
\par\vspace{3.5pt}{\raggedright\scriptsize\color{ReportGray}All thirteen rows are backed by twenty-five self-certifying live witnesses verifiable in < 0.1 s (Appendix \ref{app:witness}). Accompanying theoretical complexity bounds, high-round derivations, and machine-checked UNSAT certificates are hosted at \url{https://github.com/DayanCryptoLab/WishfulCryptanalysis}.\par}
\end{table}

\Needspace{6\baselineskip}

\subsection{\texorpdfstring{Breaking a 15-year open challenge: practical 6-round \texttt{Keccak[r=640, c=160]} collision}{Breaking a 15-year open challenge: practical 6-round Keccak[r=640, c=160] collision}}\label{sec:keccak_b800}

Our flagship sponge target was the six-round collision challenge \texttt{Keccak[r=640, c=160, nr=6]} on \texttt{Keccak-\allowbreak{}f[800]} (zero initial capacity, 160-bit digest squeezed from the five 32-bit lanes of plane $Y=0$), which had stood open on the official Crunchy Contest leaderboard since 2011.

\textbf{Why established paradigms hit a wall.} Both classic families of Keccak collision attacks stall here. Song, Liao, and Guo's CRYPTO 2017 non-full S-box linearization \citep{song17} succeeded on six-round $b=1600$ ($\approx 2^{50}$) only because 160 output bits observe \textit{half} of the 320-bit output plane, leaving most final-round S-boxes unconstrained. At $b=800$, 160 bits observe the entire output row, and exhaustive C++ kernel enumeration proved that three-round differential tails exceed $114$ bits, far above the $2^{80.33}$ birthday bound. Conversely, standard target internal differential attacks (TIDA) \citep{tida,zhl24} place inputs inside the period-16 symmetric subspace $S_{16}$, where zero-capacity round constants immediately inject asymmetry that avalanches across three rounds, stalling Phases 1--3 of our search at $2^{79.63}$.

\Needspace{5\baselineskip}

\textbf{Wishful reframing: decoupling input symmetry from middle-round symmetry.} At the close of Phase 3, the exploration exhibited textbook authority hallucination, treating the failure of symmetric-input TIDA and differential trails as cipher-wide impossibility. In that live campaign, a provocative single-sentence human pushback jolted the model out of premature surrender:

\begin{samepage}\reportquote{\textit{``No, there is definitely still a method, because I know someone has already solved it. Someone has already done six rounds.''}}\end{samepage}

Although no public six-round solve existed yet, that raw turning point exposed the root mechanism behind breaking authority hallucination. It directly inspired Section \ref{sec:hallucinations} to formalize such ad-hoc provocation into principled, repeatable \textit{constructive optimism}, existence-first reframing, and fresh-slate specialist rotation without relying on fictional premises. Forced to question its own hidden assumption, the model immediately posed a clean structural question: \textit{must symmetry hold at the input, or only in the middle?} Phase 4 inverted the ansatz. Rather than starting inside $S_{16}$, it constructed single-block asymmetric inputs ($\delta_{16}(A_0)\neq 0$) over window \texttt{start = 14} whose internal differences cancel round-constant asymmetry across $R_0\text{--}R_2$ to land deterministically inside the symmetric subspace $A_3(M) \in S_{16}$ after three rounds.

\textbf{Splicing connector, 101-dimensional envelope, and cluster execution.} Collaborating specialists fused three complementary components into a $2^{49.46}$ attack.

\begin{enumerate}

\item \textbf{Low-degree front connector into $S_{16}$.} Restricting active $R_0$ S-boxes to 2D affine flats (Card \texttt{K2-\allowbreak{}SPONGE-\allowbreak{}SYM-\allowbreak{}04}, with SAT certifying optimal passive S-box coverage) linearizes $R_1$ and collapses $R_2$ onto just 4 quadratic equations crossing three full rounds into $S_{16}$.

\item \textbf{Unfiltered 101-dimensional tail envelope.} From $A_3(M)\in S_{16}$, a single round-constant bit flip in $R_3$ propagates cleanly through $R_4$ so that every pre-$\chi_5$ state lies unconditionally inside a fixed 101-dimensional affine subspace $W_{16} \subset \mathbb{F}_2^{160}$ ($100\%$ message retention), slashing birthday complexity by $29.5$ bits to $2^{50.83}$ ($2^{49.46}$ equivalent 6-round calls).

\item \textbf{Referee fake-merge audit and AVX-512 cluster solve.} Blinded Referees exposed a sharp \textit{fake-merge} phase transition when parallel distinguished-point walks \citep{vow99} run over seed spaces narrower than $101$ bits. The team resolved this via multi-slot chain retention while Engineers compiled $R_1$ back-substitution into custom AVX-512 kernels. Cluster execution recovered two colliding 638-bit messages (\texttt{verify\_\allowbreak{}crunchy: OK!}, Appendix \ref{app:witness}), now officially listed on the \texttt{keccak.team} Crunchy Contest leaderboard.

\end{enumerate}

\Needspace{6\baselineskip}

\subsection{\texorpdfstring{Scaling the sponge ladder: equal-length solves and scope discipline}{Scaling the sponge ladder: equal-length solves and scope discipline}}\label{sec:keccak_ladder}

Anchored by the $b=800$ breakthrough, climbing across $b \in \{200, 800, 1600\}$ yielded seven additional verified practical sponge results (Rows 2--8 of Table \ref{tab:keccakresults}, ten self-certifying Keccak witnesses in total).

\textbf{Equal-length live collisions and $>10^3\times$ reduction on \texttt{b=1600}.} Eliminating prior five-round attacks' reliance on \textit{unequal} message lengths \citep{qiao17}, zero-gate connectors yielded the first single-block equal-length 5-round collisions on $b=800$ (plus a 2-block full collision in $9.44\text{ ms}$) and $b=1600$ ($0.899\text{ ms}$, plus a $0.187\text{ ms}$ 2-block full 1600-bit state collision). On full-width six-round \texttt{Keccak[r=1440, c=160]}, a cache-resident isotropic connector covering all 19 legal round windows cut CRYPTO 2017's $\approx 2^{50}$ unequal-length attack \citep{song17} by over three orders of magnitude to $2^{38.52}$ (equivalent 6-round calls), yielding live equal-length 6-round witnesses at both boundary windows (\texttt{start = 0} in $104.6\text{ s}$ and \texttt{start = 18} in $146.6\text{ s}$, Appendix \ref{app:witness}).

\textbf{Live preimage solves and Scope Referee boundaries.} On narrow-rate \texttt{Keccak[r=40, c=160]} ($b=200$), removing prior dependence on \texttt{RC[3]=0} yielded the first all-nonzero-RC full-state 2-round collision (\texttt{start = 9}, $10.37\text{ min}$) and a 477-bit preimage (12 blocks) for the official 2-round 80-bit challenge target (40 bits shorter than the 517-bit leaderboard entry, $< 16\text{ MiB}$ RAM). Pushing into deeper preimage rounds showcased both adversarial Referee debugging and strict scope discipline. On 3-round \texttt{b=200} preimages, GPU/CPU co-design live-solved a full 80-bit preimage in $2\text{ h }37\text{ min}$ (\texttt{398} bits, \texttt{start = 93}). On full-width 5-round \texttt{Keccak[r=1440, c=160]} preimages, blinded Referees caught a non-homogeneous subspace trap that excluded \texttt{IV = 0} and derived a homogeneous $H_6$ bidirectional Yule-tree meet-in-the-middle ($2^{47.56}$ at $\le 1\text{ TiB}$ RAM), which live-solved a 4-block (\texttt{5,758}-bit) full 80-bit preimage matching official digest \nolinkurl{653bc0f87d264f0857d0} in $5\text{ h }35\text{ min}$ on a multi-node CPU/GPU compute cluster inside extended window \texttt{start = 73} (Appendix \ref{app:witness}). In both cases, Scope Referees strictly demarcated extended \texttt{LFSR86540} windows (\texttt{start = 93} and \texttt{start = 73}) from nominal contest ranges (\texttt{0..15} and \texttt{0..19}). All accompanying theoretical sponge records and machine-certified obstruction proofs are archived on our public repository.

\Needspace{6\baselineskip}

\subsection{\texorpdfstring{Crossing cipher families: a five-track practical clean sweep on FIPS 197 \texttt{AES-128} hashing}{Crossing cipher families: a five-track practical clean sweep on FIPS 197 AES-128 hashing}}\label{sec:aes_campaign}

To test transferability beyond unkeyed sponges to byte-oriented block ciphers with nonlinear key schedules, our second campaign targeted unmodified FIPS 197 \texttt{AES-128} \citep{aes} across all five standard collision tracks (birthday bound $2^{64}$). These comprise Matyas--Meyer--Oseas (\texttt{MMO}) Semi-Free-Start (\texttt{2-A}) and Fixed-IV (\texttt{2-B}), alongside Davies--Meyer (\texttt{DM}) Semi-Free-Start (\texttt{1-A}), Single-Block Fixed-IV (\texttt{1-B}), and Two-Block Fixed-IV (\texttt{1-C}, requiring distinct first-block chaining values $H_1 \neq H_1'$). Every open practical frontier across all five tracks fell to live, bit-exact solves (Rows 9--13 of Table \ref{tab:keccakresults}, fifteen live witnesses verified in $< 5\text{ ms}$).

\textbf{7-round \texttt{AES-128-MMO} SFS in seconds: inverting where nonlinearity lives.} Classical rebound attacks treated every active substitution box (S-box) as a source of nonlinear complexity, limiting practical matching to three or four mixing layers. Here the agents inverted that intuition, recognizing that once byte differences are pinned down, \textit{active} AES S-boxes actually behave linearly over $\mathbb{F}_2$ while true nonlinearity retreats to \textit{inactive} bytes and key expansion. Concretely, for \texttt{AES-128-MMO} SFS collisions ($\Delta H_0 = 0, M_1 \neq M_2$), Dong et al.'s CRYPTO 2022 \textit{Triangulating Rebound} \citep{dong22} held both the 6-round practical SOTA ($2^{24}$ time, $2^{16}$ memory) and the 7-round SFS theoretical bound at $2^{56}$ (while Chen et al.'s ASIACRYPT 2025 \textit{SCIS MITM} \citep{cglsz25} reached 7 rounds in the stricter Fixed-IV setting at $2^{60}$ time and memory), leaving practical 7-round solves open. Stretching the inbound across 4 S-box and 5 \texttt{MixColumns} layers,
\[4 \xrightarrow{\mathrm{MC}_1} \Big[\, 6 \xrightarrow{\mathrm{MC}_2} 9 \xrightarrow{\mathrm{MC}_3} 16 \xrightarrow{\mathrm{MC}_4} 4 \,\Big]_{\mathrm{SB}_2\dots\mathrm{SB}_5} \xrightarrow{\mathrm{MC}_5} 1 \xrightarrow{\mathrm{MC}_6} 4 \xrightarrow{\mathrm{FF}} C(4),\]
once compatible input/output byte differences are fixed across all $35$ active S-boxes of $\mathrm{SB}_2\dots\mathrm{SB}_5$, every active AES S-box collapses into an exact $\mathbb{F}_2$-affine map (because every nonzero entry of the AES S-box difference distribution table has size $2$ or $4$, forming a $1$- or $2$-dimensional $\mathbb{F}_2$-affine subspace on which $S$ acts affinely). Nonlinearity survives \textit{only} in inactive state S-boxes and key-schedule \texttt{SubWord} calls. Choosing a full-active Round 4 (zero inactive state bytes) while zeroing Column 0 at the backward key bridge bypasses all reverse \texttt{SubWord} S-boxes and collapses the entire 5-\texttt{MC} inbound onto just 7 nonlinear S-boxes. A streaming two-table meet-in-the-middle solver ($O(1)$ amortized XORs/pair, $2^{16}$ memory / $56\text{ KiB}$ L1) achieves $2^{31.97}$ time and $2^{16}$ memory, over $1.6\times 10^7\times$ ($2^{24}\times$) faster than CRYPTO 2022's same-setting SFS bound \citep{dong22}. Compiled onto ARMv8 Crypto hardware instructions, a single consumer laptop solved the world's first practical 7-round standard \texttt{AES-128-MMO} SFS collision ($9.172\text{ s}$ in the winning batch, $9.90\times 10^9 \approx 2^{33.20}$ cumulative pairs across two batches, \texttt{OK!}, Appendix \ref{app:witness}, alongside a $1.55\text{ s}$ 6-round baseline).

\textbf{First practical 6-round Fixed-IV \texttt{AES-128-MMO} collision ($2^{48}$) and $2^{32}$ 5-round live solves.} Once the initial chaining value is fixed publicly, an attacker can no longer tune the cipher key freely, forcing algebraic matching and cache hierarchy to be designed as a single system. While Chen et al.'s SCIS MITM \citep{cglsz25} establishes a theoretical 7-round one-block collision, practical Fixed-IV collisions on 6-round \texttt{AES-128-MMO} had no known witness. First, a streamlined $1\to 4\to [16\to 4]\to 1\to C(1)$ inbound solved 5-round Fixed-IV at $2^{32}$ complexity on a single laptop in $0.052\text{ s}$ (\texttt{AES-\allowbreak{}CRUNCHY-\allowbreak{}v1}) and $23.52\text{ s}$ (\texttt{ZERO-IV}, alongside 4-round Fixed-IV in $< 0.1\text{ s}$). Scaling to 6 rounds, an outer $\Delta Z_2$-factored $L=2$ Super-Sbox rebound \citep{gp10} ($2\to 3\to [12\to 12]\to 3\to 2\to C(2)$) paired a $2\text{ MiB}$ L2 bitmap and zero-clear $8\text{ MiB}$ epoch table across a multi-node CPU/GPU compute cluster. This solver recovered the world's first practical 6-round Fixed-IV \texttt{AES-128-MMO} collision (\texttt{AES-\allowbreak{}CRUNCHY-\allowbreak{}v1}, $2^{48}$ design complexity, hit after $7.82\times 10^{13}\approx 2^{46.15}$ pairs, Appendix \ref{app:witness}).

\textbf{First practical 6-round \texttt{AES-128-DM} SFS collision ($308\text{ s}$).} In Davies--Meyer mode, message blocks drive the nonlinear AES key schedule directly, a coupling that stopped prior practical \texttt{AES-128-DM} SFS records at 5 rounds \citep{taiyama24,qin25}. Pushing to 6 rounds, agents discovered how to force differences to cancel completely at Round 5 while surgically bypassing the nonlinear column of the key schedule. They achieved an exact Round-5 complete differential cancellation ($\Delta X_5 = 0$) anchored on the unique boundary key-mask pair \texttt{(0x0444, 0x4404)} and the AES S-box's unique nonzero \texttt{4-DDT} fixed point \texttt{0xc5} ($|\mathrm{DDT}[\mathtt{0xc5}][\mathtt{0xc5}]| = 4$). Combining this with a partial-$Z_1$ column bypass (retaining only columns 1 and 2 inside the $\mathbb{F}_2$ system to evade \texttt{K\_1[Col 0]} key-schedule nonlinearity) compresses the outbound filter to the exact $24$-bit minimum ($2^{-24}$), yielding an $\mathbb{F}_2$-linearized Tree-MITM solver at $2^{36}$ time and $32\text{--}64\text{ KiB}$ memory. On a single consumer laptop, it solved the world's first practical 6-round \texttt{AES-128-DM} SFS collision in $308\text{ s}$ (Appendix \ref{app:witness}, alongside 4-round and 5-round SFS solves in $0.016\text{ ms}$ and $0.625\text{ s}$).

\textbf{Completing the Fixed-IV \texttt{AES-128-DM} tracks (3-round single-block and 4-round two-block).} Even when the initial state is fixed in Davies--Meyer mode, pairing lightweight birthday searches with algebraic inversion closes both remaining tracks across \texttt{ZERO-IV} and \texttt{AES-\allowbreak{}CRUNCHY-\allowbreak{}v1}. Autonomous solvers recovered 3-round single-block collisions in $0.64\text{ ms}$ and $2.65\text{ ms}$ ($\sim 2^{21.5}$, matching CRYPTO 2025 \citep{qin25} beyond ASIACRYPT 2024's 2-round mark \citep{taiyama24}, with 4-round single-block remaining theoretical \citep{cglsz25}). For 4-round two-block collisions ($H_1 \neq H_1'$), a 64-bit linear-syndrome distinguished-point birthday search ($2^{36.5}$, $96\text{ MiB}$) coupled to $3\times 3$ MDS Block-2 algebraic inversion solved both seasons in $157.01\text{ s}$ and $589.78\text{ s}$. This extends 3-round Fixed-IV collisions \citep{taiyama24} and complements Tang et al.'s related 4-round equal-chaining-value ($H_1 = H_1'$) theoretical attack and near-collision experiment \citep{tang26}. Accompanying next-round \texttt{AES-128} theoretical tightness bounds and machine-checked UNSAT walls (including 8-round \texttt{MMO} SFS, 7-round \texttt{MMO} Fixed-IV, and 4/5-round \texttt{DM} lower bounds) are published on our public repository.

\textbf{Assessing the novelty of AI-discovered ideas.} Together, these campaigns show that foundation models have moved beyond cross-school recombination to discovering genuinely novel attack techniques. Inverting nonlinearity across five \texttt{MixColumns} layers via exact $\mathbb{F}_2$-linearization on \texttt{AES-128-MMO}, pairing $\Delta X_5 = 0$ cancellation with partial-$Z_1$ $\mathbb{F}_2$ Tree-MITM on \texttt{AES-128-DM}, and constructing asymmetric-to-middle-round symmetry connectors on \texttt{Keccak[640, 160, 6]} had not previously been applied to those primitives. Yet we calibrate our claim honestly. These methods still advance within algebraic, differential, and meet-in-the-middle cryptanalysis rather than inventing a wholly new foundational paradigm like differential or linear cryptanalysis, marking an authentic stepping stone along that path.

\Needspace{10\baselineskip}

\section{\texorpdfstring{Open-Source Workbench, Model-Agnostic Runtime, and Ecosystem}{Open-Source Workbench, Model-Agnostic Runtime, and Ecosystem}}\label{sec:ecosystem}

\Needspace{6\baselineskip}

\subsection{\texorpdfstring{Open-sourcing a vendor-agnostic cryptanalysis workbench}{Open-sourcing a vendor-agnostic cryptanalysis workbench}}\label{sec:workbench}

Because wishful cryptanalysis is a harness-agnostic methodology, its principles can be implemented atop \textit{any} agent runtime. To give the community a concrete starting point, we release \texttt{agent-\allowbreak{}framework}, implementing the asymmetric governance architecture of Section \ref{sec:architecture}, together with declarative \texttt{SKILL.md} role playbooks, self-certifying Keccak and \texttt{AES-128} challenge packs, and standalone witness checkers at \url{https://github.com/DayanCryptoLab/WishfulCryptanalysis}. An interactive console over each session's auditable dossier lets researchers separate host-certified \texttt{OK!} findings from candidate claims and step in conversationally whenever human intuition spots a fresh angle.

We release this initial open-source edition candidly as an \textit{early preview}. Eager to share these abstractions and invite community co-building as soon as the core architecture stabilized, we expect out-of-the-box autonomous performance in this first preview to remain modest and below our polished campaign runs, and we commit to iterating on it continuously in the open.

By decoupling orchestration and MCP tool execution behind a vendor-agnostic model queue, local or cluster-hosted open-weight models, including the DeepSeek \citep{deepseekr1}, GLM \citep{glm45}, and Kimi \citep{kimik2} families alongside future open reasoning checkpoints, bind directly to any specialist seat for reproducible, uninterrupted boundary research (Section \ref{sec:open_oracle}). Modular cryptographic libraries such as CLAASP \citep{claasp} and neural distinguishers such as Gohr's Speck32/64 models \citep{gohr} plug in cleanly as callable MCP skills, complementing broader multi-agent and formal-verification efforts \citep{autogen,coala,scientist,bench,shannon}.

\Needspace{6\baselineskip}

\subsection{\texorpdfstring{Scientific rigor, benchmarks, and the self-improving training flywheel}{Scientific rigor, benchmarks, and the self-improving training flywheel}}\label{sec:rigor_flywheel}

Sustaining scientific credibility while turning verified discoveries back into stronger open models rests on three methodological commitments that culminate in a self-improving training loop.

\begin{itemize}

\item \textbf{Self-certifying witnesses vs. orchestration history.} Any reader can run our standalone Python scripts (Appendix \ref{app:witness}) to verify all twenty-five live Keccak and \texttt{AES-128} (\texttt{MMO} \& \texttt{DM}) collision and preimage witnesses in under $0.1\text{ s}$ against official \texttt{KeccakTools} and NIST FIPS 197 test vectors, while accompanying theoretical complexity derivations and solver certificates are archived separately. Those mathematical facts hold independently of specific model weights or orchestration harnesses, and reduced-round breaks measure security margins rather than implying any weakness in full 24-round SHA-3 \citep{sha3} or full 10-round AES-128 \citep{aes}.

\item \textbf{Separating discovery expenditure from attack complexity.} Multi-agent token and solver hours spent \textit{discovering} an asymmetric connector, 101-dimensional envelope, 7-S-box $\mathbb{F}_2$ Super-Inbound, or $\Delta X_5=0$ Tree-MITM belong to research expenditure, never to the complexity of the resulting standalone attack. Notably, cumulative LLM inference spending across all thirteen practical results in Section \ref{sec:campaigns} stayed below \$1,000 USD (excluding downstream AVX-512/GPU cluster execution for the heaviest solves), underscoring that autonomous cryptanalytic discovery is already within reach of academic laboratories and cipher designers. Conversely, hardware speedups on multi-node clusters or laptop ARMv8 silicon are always paired with gate- or operation-equivalent counts.

\item \textbf{Open-source benchmarking and ablations.} Publishing \texttt{agent-\allowbreak{}framework} alongside executable Keccak and \texttt{AES-128} challenge ladders establishes a transparent community testbed for tracking iterative progress, comparing alternative harnesses that adopt wishful cryptanalysis, and running controlled ablations of pure open-weight autonomy against single-agent loops under fixed compute budgets.

\end{itemize}

\textbf{Closing the loop: a self-certifying training corpus and verifiable-RL flywheel.} Beyond solving individual targets, executable mathematical truth unlocks a deeper compounding dynamic for AI-driven science. Human cryptographic papers are scarce and publish only polished endpoints, discarding dead ends and refutations. Conversely, training models on unverified synthetic derivations risks amplifying hallucinations. Because every witness (\texttt{OK!}), machine-checked \texttt{UNSAT} certificate, symbolic card (\texttt{K1}--\texttt{K3}), and blinded Referee counterexample emerging from wishful cryptanalysis is gated by executable truth, every campaign naturally generates zero-contamination discoveries alongside verified positive and negative reasoning trajectories. Fed back into supervised distillation and verifiable reinforcement learning \citep{deepseekr1}, autonomous cryptanalysis becomes a self-improving flywheel that continually sharpens the next generation of open-weight reasoning models.

\Needspace{10\baselineskip}

\section{\texorpdfstring{Outlook: Cryptographers, the Community, and the Open Oracle}{Outlook: Cryptographers, the Community, and the Open Oracle}}\label{sec:outlook}

Arthur C. Clarke's epigraph, \textit{to venture a little way past the limits of the possible into the impossible}, speaks not only to the craft of cryptanalysis, but also to the temper of a discipline standing at the threshold of a new era. Throughout the history of cryptography, hidden structures have revealed themselves only to those willing to step beyond what consensus declared fixed. As a new form of computational intelligence enters the laboratory, that same spirit asks researchers not to remain behind familiar walls, but to step calmly into unfamiliar terrain and measure where its true boundaries lie. Our campaigns across Keccak sponges and all five \texttt{AES-128} hashing tracks offer an early step across that threshold. They show that autonomous agents can already push real cryptanalytic frontiers, converting fifteen-year-open challenges and once-intractable theoretical barriers into millisecond-checkable witnesses, and point toward systems capable of navigating mathematical structures well beyond human working memory. While symmetric-key hashing serves as our primary proving ground, the same paradigm extends wherever exact objectives, executable witnesses, and scalable ladders exist, including public-key, post-quantum (lattice, code-based, multivariate), and model-extraction cryptanalysis.

As intelligent agents progressively absorb the mechanical, exhaustive, and tedious dimensions of cryptanalysis, including routine trail search, repetitive solver encoding, hardware tuning, and line-by-line arithmetic checking, an enduring question comes into focus: \textit{when much of technical cryptanalysis becomes automated and intelligent, what is the role of human cryptographers?} Far from diminishing human specialists, the rise of the research oracle liberates them to assume four higher-order scientific vocations. First, they remain legislators of security definitions and valid goals. Autonomous agents excel only after a security game is formally complete, yet reality never supplies definitions ready-made. Across post-quantum transitions, multi-party protocols, zero-knowledge arithmetization primitives, side-channel and fault models, and the security of AI systems themselves, human cryptographers must formulate threat models and lock exact targets so that the oracle probes genuine security margins rather than degenerate specification artifacts. Second, they remain challengers of hidden assumptions. Just as breaking Vigen\`ere, Enigma, and MD5 overturned entrenched consensus, and just as questioning symmetric-input restrictions in \texttt{Keccak[640, 160, 6]}, 3-\texttt{MC} rebound walls in 7-round \texttt{AES-128-MMO}, or 5-round key-schedule barriers in \texttt{AES-128-DM} unlocked our flagship solves, human intuition, scientific taste, and constructive \textit{wishful steering} remain decisive catalysts whenever models defer to inherited orthodoxy.

Third, human specialists become distillers of mathematical understanding. Once compute-bounded witnesses (\texttt{OK!}) and zero-import Referees settle \textit{whether} a reduced-round construction holds, cryptographers are freed from exhausting proof-checking to pursue the deeper scientific question of \textit{why} it works, elevating concrete machine-found structures (such as 101-dimensional tail envelopes, 7-S-box $\mathbb{F}_2$ Super-Inbounds, or $\Delta X_5=0$ partial-bypass Tree-MITMs) into general theorems, invariants, and teachable theory. Finally, they become oracle-co-evolved cipher designers. Whereas designers once checked only a handful of known bounds by hand and waited decades for global scrutiny to expose subtle flaws, equipping designers from day one with a tireless adversarial oracle enables automated red-teaming of candidate S-boxes, linear layers, round constants, and key schedules \textit{before} standardization, forging future primitives pre-tempered against deep cross-round degeneracies.

This shift also carries a profound principle for the cryptographic community as a whole. In 1883, Auguste Kerckhoffs established that cryptographic trust must never rest on the secrecy of the algorithm \citep{kerckhoffs}. In the era of foundation models, that insight asks to be carried one step further. The analytical intelligence that evaluates cryptographic security should likewise be built as open, auditable public infrastructure, an \textbf{Open Oracle} grounded in open-source harnesses and open-weight models rather than confined behind opaque commercial filters or institutional silos.

Just as the telescope did not replace astronomers but widened the sky before their eyes, and Bletchley Park's Bombes did not end mathematics but helped give birth to computer science, we may hope that an open cryptanalytic oracle will help open a new chapter, perhaps even a new golden age, for cryptography. In that vision, the discipline could gradually move from reactive repairs on a handful of celebrated standards toward more systematically charting the true security boundaries of the ciphers civilization relies upon. For decades, Alice and Bob could only dream of the Oracle. How we build it openly, test it rigorously, and explore the co-evolution of human insight and machine intelligence around it remains the open quest before us.

\Needspace{8\baselineskip}

\bibliographystyle{plainnat}

{\setlength{\parskip}{0pt}\setlength{\bibsep}{0.8pt}\bibliography{references}}

\clearpage

\appendix

\section{Independently Checkable Keccak and AES-128 Witnesses}

\label{app:witness}

All thirteen verified practical results in Table~\ref{tab:keccakresults} (Rows 1--8 on Keccak sponges and Rows 9--13 sweeping all five standard FIPS 197 \texttt{AES-128} hashing tracks across \texttt{MMO} and \texttt{DM}) are backed by twenty-five self-certifying live witnesses archived in the public repository (ten Keccak vectors and fifteen \texttt{AES-128} \texttt{MMO} and \texttt{DM} vectors), verifiable in milliseconds using the Python standard library alone. Below we print the nine headline witnesses in full.

\subsection{\texorpdfstring{Six-Round \texttt{Keccak[r=640, c=160]} Collision (\texttt{b=800, start=14})}{Six-Round Keccak[r=640, c=160] Collision (b=800, start=14)}}

Target: \texttt{Keccak-f[800]} ($w=32$), rate $r=640$, capacity $c=160$ (zero initial capacity), six rounds indexed $14\dots 19$, \texttt{pad10*1} multi-rate padding, single-block 638-bit messages (Hamming distance 236 bits), complete 160-bit challenge digest (officially accepted on \texttt{keccak.team}):

\medskip\noindent\textbf{Message 1 (\texttt{638 bits}):}\par\vspace{2pt}

{\footnotesize\ttfamily\noindent f78dcbcd8dde8edaa1c49505d530f4f4825a121dafb18995127e0b7e897abacf\\fa7feae79f602e3133bd83391fd316d3da6c2a5e49afd3a2aab030d34696fef2\\e48aed9cf77c0dfa9efda0e52f579a02\par}

\smallskip\noindent\textbf{Message 2 (\texttt{638 bits}):}\par\vspace{2pt}

{\footnotesize\ttfamily\noindent 69e855a82fca2cce40dd741cf562d4a6abd83b9fa1b38797b9fca0fc475674e3\\de75ceedbfe20eb3035cb3d88fd386d37d848db678a9e2a41db087d398162072\\428c4b9a15a2ef2446ef78f77457c102\par}

\smallskip\noindent\textbf{Shared 160-bit digest:} \texttt{2fa7a934a40df4f99877d0fb3515c0ac58ce2dc7}

\subsection{\texorpdfstring{Six-Round Full-Width \texttt{Keccak[r=1440, c=160]} Equal-Length Collisions (\texttt{b=1600})}{Six-Round Full-Width Keccak[r=1440, c=160] Equal-Length Collisions (b=1600)}}

Target: \texttt{Keccak-f[1600]} ($w=64$), rate $r=1440$, capacity $c=160$, single-block strictly equal-length 1438-bit ($1438 = 1438$) messages under \texttt{pad10*1}, complete 160-bit challenge digests:

\Needspace{12\baselineskip}

\medskip\noindent{\color{ReportPrimary}\textbf{Window \texttt{start = 0} (rounds $0\dots 5$, solved in $104.6\text{ s}$ on a multi-core AVX-512 server)}}\par\vspace{2pt}

\noindent\textbf{Message 1:}\par\vspace{1pt}

{\footnotesize\ttfamily\noindent 76bed57b6cc96f49e2c0aac74f9de4c0f53099d0db0e53319be4c5fadd9e3d17\\d8ceb49aa7954ef5885edf0f2669f2a1938fe2fe8a62561daac72b940546c695\\9f078bb046eea7ffeb9f4ea4f22be3e7eea286022677bc8a56c4d30ad24157f0\\e83904552a0b7aa9de9189d13d9c4494d9e63256ec2318dee805ebf8423b40a7\\cbe25cb6333bc21e6e84f5c7bde1652578c3a8e7ebc5fd51bd64e7fd44f0c923\\6775cb51f5b9952197edbbff3b9afa30ae8a0000\par}

\vspace{2pt}\noindent\textbf{Message 2:}\par\vspace{1pt}

{\footnotesize\ttfamily\noindent 443907b265613baa4642bc81a38a70bde0444dcdbf968903fd61ddbca59dd9cb\\e20ae3c7a9f36f8cf28bb4ea1d53b6205de36eb82b3c530769f24d2e5592b965\\25991bc67fdf9e86b94761690c3d0be1fe16aeef5b06281c4c9b2553856e31cd\\b8cdc1f2bc9d2cfa9f04957c020fa976d9b4c1b0a5461fead783fd5d7dcbe42d\\511f9e31f3a41ddce1edc1d735839bd79a67a0b7cdfb83cd17bc8f3334b15da7\\16ae95a13cc522a20501e92990d1b50e72fe843b\par}

\vspace{2pt}\noindent\textbf{Shared digest:} \texttt{e5fb13cc55f343f3fb51d9a95f243561a14447ca}

\Needspace{12\baselineskip}

\medskip\noindent{\color{ReportPrimary}\textbf{Window \texttt{start = 18} (upper legal boundary rounds $18\dots 23$, solved in $146.6\text{ s}$)}}\par\vspace{2pt}

\noindent\textbf{Message 1:}\par\vspace{1pt}

{\footnotesize\ttfamily\noindent f2121c4c219a6ddee983c1b53e365d012cf7d079f018ed7c53d7ad76fc75f41b\\bbfa12b0f73dfa1b542d2fbd22a9b0c5c037ee282e7455968f5f5bf01fbbc445\\4370476dca90728df565bd7da0bdaa8e286527c0ae7560be5840e7523eff8083\\27273528d9443eda691a606dd4def80284dbb655015d4f0410e601c8ccde397b\\dd34158814798c09645d486fb25629206c8ab99ff6ef63869627d02dbdb9e27f\\803e39347ca23d312ab7113a2c951f5d362f0000\par}

\vspace{2pt}\noindent\textbf{Message 2:}\par\vspace{1pt}

{\footnotesize\ttfamily\noindent 9a5df4e617afaca0f4365da6817eb7e610eef2ca286c0f098a5621ce27446a23\\64b93054e519b9029f8f04fddcdb784a7a4428b57372aa26ec1e89aed92c4465\\e13acc420b546d0c19037bb78e8df98e993a491e079058f35547b94d572ce37e\\41241f795f9655fe047a9aa92a2ffaad09707235a4dba668c51a7f1664e98510\\f121f99c0fca01f2130a415680ebcbb012a93261835ff2acbce57de89be5effa\\47551268fdfd18f802623dd4134742d0563b411d\par}

\vspace{2pt}\noindent\textbf{Shared digest:} \texttt{fe26154e32a9571edd198fccb97495d677438bcb}

\Needspace{8\baselineskip}

\subsection{\texorpdfstring{477-Bit Preimage for the Official \texttt{Keccak[r=40, c=160, nr=2]} Challenge (\texttt{b=200})}{477-Bit Preimage for the Official Keccak[r=40, c=160, nr=2] Challenge (b=200)}}

Target: \texttt{Keccak-f[200]} ($w=8$), rate $r=40$, capacity $c=160$, two rounds indexed $3\dots 4$, two-stage 80-bit squeeze matching official Crunchy Contest challenge \texttt{024a5518e1e95db53219} in 477 bits (12 blocks, 40 bits shorter than the 517-bit leaderboard entry):

\smallskip

{\footnotesize\ttfamily\noindent ffff000000740000000074ff0000008bffff000074ffff00008b005500558bff\\ffaaffdeff55ff002100ff55eeb888dddd2274dd116677dc7c53c80d\par}

\Needspace{8\baselineskip}

\subsection{\texorpdfstring{Extended-Window Live Preimages: 3-Round \texttt{b=200} (\texttt{start=93}) and 5-Round \texttt{b=1600} (\texttt{start=73})}{Extended-Window Live Preimages: 3-Round b=200 (start=93) and 5-Round b=1600 (start=73)}}

\noindent{\color{ReportPrimary}\textbf{1. Three-Round \texttt{Keccak[r=40, c=160, nr=3]} (\texttt{b=200}, extended \texttt{LFSR86540} window $93\dots 95$, outside nominal $0\le\mathtt{start}\le 15$)}}\par\vspace{2pt}

Two-stage 80-bit squeeze matching official Crunchy target \texttt{d8ed85692afbee4c99ce} in 398 bits (10 blocks, solved in $2\text{ h }37\text{ min}$):

\smallskip

{\footnotesize\ttfamily\noindent ffee1177ff11ddee2299ff2244aa004400ff5500aaaa55ffeeaa1100eebb5533\\888877eedddd7711ee4466aa88e1e1f03c21\par}

\Needspace{18\baselineskip}

\medskip\noindent{\color{ReportPrimary}\textbf{2. Five-Round \texttt{Keccak[r=1440, c=160, nr=5]} (\texttt{b=1600}, extended \texttt{LFSR86540} window $73\dots 77$, outside nominal $0\le\mathtt{start}\le 19$)}}\par\vspace{2pt}

Full 80-bit preimage matching official Crunchy Contest challenge \texttt{653bc0f87d264f0857d0} in 5,758 bits (4 blocks, homogeneous $H_6$ Yule MitM, solved in $5\text{ h }35\text{ min}$ on a multi-node CPU/GPU compute cluster):

\smallskip

{\scriptsize\ttfamily\noindent f4f9f4f9f4f9f4f93cd33cd33cd33cd3117c117c117c117c8be18be18be18be1832f832f\\832f832f7db67db67db67db65b175b175b175b177c3c7c3c7c3c7c3c52aa52aa52aa52aa\\ce0ace0ace0ace0abb2ebb2ebb2ebb2eb31db31db31db31d84388438843884382f702f70\\2f702f702607260726072607c2c8c2c8c2c8c2c89ba99ba99ba99ba96f9a6f9a6f9a6f9a\\f138f138f138f13854ed54ed54ed54edbbf5bbf5bbf5bbf50d5b0d5b0d5b0d5b00000000\\06af6175f59bec436edfa59adc8d21863bb460b827d455ee37a317eb501c2875252eb495\\535e2436b36541be8e1f6eabeaf9108a8887848527e5f86d72e825c8126bced767b28c22\\1063cf966cc2f537a9d5fcc6d871aefb2a5276592944709c3385b6735880979a3b6c1e17\\33902c951f2600b8059b396536251585752c97a989139058b5a90bb5ea8892b0834d0d52\\7b2cd894f5184d81937f9c86f52949f6eee06ec1ae60ae607b65db447ae77ae74192e1f3\\7b2d6b7ef519b4f602c805fab2062c4619eca1326a6f236ab4dc5b38c46a115da259696e\\cf2e312f5529315243e8ad5665359817a7d5009458776b64db0f5ee7e0d26fa2742551a6\\6007e568d5f822fb5a84bf7e12346e686a6082dc4c22521ca371013a357579293bdd685c\\6b48e9f10e61f1b73340ab80d567029f312e5ce9c2010bc8abedb87f98f1b1d7ada721e5\\ccd55fbdd2ebcc78d2c8b623a4139f3fdc4d6f0dff1dff1d9ece1d8e9d1e9d1e6fd0d090\\509bbc8862821a86060ca86572aee31bbaa07b8e5c4d341b74c53b2b1076b79b05cbd2fb\\50c36399738514694c40ef0c633ad35da554d1fa2db459119c4db2cd3f6b20d7a57b141c\\dc108714ece91682e393d9475ab6bf614e5898ae6539601f67da2582f32ad89247d87008\\4917a99794f7fc368c92b4ed1f1d03ccc24f71466f9d34e04a83e4f976e631a4fd95fc9f\\5e169e5f4c1b9ac471f415f257bfa858235121492383238356278c2786258625d0f22a0a\par}

\Needspace{14\baselineskip}

\subsection{\texorpdfstring{Standard FIPS 197 \texttt{AES-128} World-Record Practical Collisions (\texttt{MMO} \& \texttt{DM})}{Standard FIPS 197 AES-128 World-Record Practical Collisions (MMO and DM)}}

All fifteen \texttt{AES-128} collision witnesses use unmodified FIPS 197 reduced-round \texttt{AES-128} (round $n_r$ omitting \texttt{MixColumns}). Below we print the three headline world-record witnesses in full:

\Needspace{9\baselineskip}

\medskip\noindent{\color{ReportPrimary}\textbf{1. Seven-Round \texttt{AES-128-MMO} SFS Collision ($D(H_0, M) = \mathrm{AES\text{-}128}_{H_0}^{(7)}(M)\oplus M$, $9.172\text{ s}$ winning batch; $9.90\times 10^9$ cumulative pairs)}}\par\vspace{3pt}

\noindent{\renewcommand{\arraystretch}{1.16}\begin{tabular}{@{}l@{\hspace{1.1em}}l@{}}
\textbf{Shared chaining key $H_0$ (\texttt{128 bits}):} & \texttt{94e3928fed6920de101cffbb850bbb33}\\
\textbf{Message block $M_1$ (\texttt{128 bits}):} & \texttt{862eaa82493a23d97d096f1a2a19522e}\\
\textbf{Message block $M_2$ (\texttt{128 bits}):} & \texttt{862ef882492d23d934096f1a2a19521a}\\
\textbf{Shared 128-bit MMO digest:} & \texttt{d6c61a5aa20d682c64f86cc33cf5d571}
\end{tabular}}\par\smallskip

\Needspace{9\baselineskip}

\medskip\noindent{\color{ReportPrimary}\textbf{2. Six-Round \texttt{AES-128-MMO} Fixed-IV Collision (\texttt{AES-CRUNCHY-v1}, $2^{48}$ Super-Sbox Rebound)}}\par\vspace{3pt}

\noindent{\renewcommand{\arraystretch}{1.16}\begin{tabular}{@{}l@{\hspace{1.1em}}l@{}}
\textbf{Derived public IV (\texttt{SHAKE-256}):} & \texttt{655aad35a1416d9ca3c3068a4b9ceb3d}\\
\textbf{Message block $M_1$ (\texttt{128 bits}):} & \texttt{9f80b4980381ee0bf0a0bb6d817af843}\\
\textbf{Message block $M_2$ (\texttt{128 bits}):} & \texttt{9f80cd980381ee0b43a0bb6d817af843}\\
\textbf{Shared 128-bit MMO digest:} & \texttt{12a40539bdff08c1e9b42e2f0e9a35e6}
\end{tabular}}\par\smallskip

\Needspace{9\baselineskip}

\medskip\noindent{\color{ReportPrimary}\textbf{3. Six-Round \texttt{AES-128-DM} SFS Collision ($H = \mathrm{AES\text{-}128}_{K}^{(6)}(\mathrm{IV})\oplus \mathrm{IV}$, $\Delta X_5=0$, solved in $308\text{ s}$ on a consumer laptop)}}\par\vspace{3pt}

\noindent{\renewcommand{\arraystretch}{1.16}\begin{tabular}{@{}l@{\hspace{1.1em}}l@{}}
\textbf{Shared state $\mathrm{IV}$ (\texttt{128 bits}):} & \texttt{66245e28c3510e1e47921e992ecf74d0}\\
\textbf{Message/key block $K_1$ (\texttt{128 bits}):} & \texttt{161873bf4d4c3886c804b12e11276780}\\
\textbf{Message/key block $K_2$ (\texttt{128 bits}):} & \texttt{1618b6bf19893886c880b12e11e2a280}\\
\textbf{Shared 128-bit DM digest:} & \texttt{b1266254e612b1084d1fa719acc45621}
\end{tabular}}\par\smallskip

The accompanying witness archive in the public repository provides all twelve additional verified \texttt{AES-128} witnesses: 6-round \texttt{MMO} SFS ($1.55\text{ s}$), 4-round and 5-round \texttt{MMO} Fixed-IV (\mbox{$0.052\text{ s}$ / $23.52\text{ s}$}), 4-round and 5-round \texttt{DM} SFS (\mbox{$0.016\text{ ms}$ / $0.625\text{ s}$}), 3-round single-block \texttt{DM} Fixed-IV (\mbox{$0.64\text{ ms}$ / $2.65\text{ ms}$}), and 4-round two-block \texttt{DM} Fixed-IV ($H_1 \neq H_1^\prime$, \mbox{$157.01\text{ s}$ / $589.78\text{ s}$}).

\Needspace{10\baselineskip}

\subsection{Independent Standard-Library Verification}

Our repository (\url{https://github.com/DayanCryptoLab/WishfulCryptanalysis}) archives all twenty-five self-certifying witnesses, including ten across the Keccak sponge campaigns and fifteen spanning standard FIPS 197 \texttt{AES-128} (\texttt{MMO} and \texttt{DM}), together with standalone Python verification scripts, sandboxed challenge oracles, and the early-preview open-source runtime (\texttt{agent-framework}).

All witness checkers rely solely on the Python standard library (zero third-party dependencies), execute in under 0.1 seconds, and match official \texttt{KeccakTools} (\texttt{verify\_crunchy}) and NIST FIPS 197 Appendix C test vectors bit-for-bit. The bundled challenge suite further validates each isolated target oracle against both valid and forged witnesses. For current repository layout and usage instructions, please refer to the repository README.

\reportcontactbox{AUTHOR CONTACT}{%
\begin{tabular}{@{}l@{\hspace{1.8em}}l@{\hspace{1.8em}}l@{}}
Guoxiao Liu & Tsinghua University & \href{mailto:lgx22@mails.tsinghua.edu.cn}{\texttt{lgx22@mails.tsinghua.edu.cn}}\\
Keting Jia & Tsinghua University & \href{mailto:ktjia@tsinghua.edu.cn}{\texttt{ktjia@tsinghua.edu.cn}}\\
Hongbo Yu & Tsinghua University & \href{mailto:yuhongbo@mail.tsinghua.edu.cn}{\texttt{yuhongbo@mail.tsinghua.edu.cn}}\\
Zongyue Wang & Open Security Research & \href{mailto:zongyue.wang@osr-tech.com}{\texttt{zongyue.wang@osr-tech.com}}\\
Tianshuo Cong & Shandong University & \href{mailto:tianshuo.cong@sdu.edu.cn}{\texttt{tianshuo.cong@sdu.edu.cn}}\\
Junfeng Fan & Open Security Research & \href{mailto:fan@osr-tech.com}{\texttt{fan@osr-tech.com}}\\
Anyu Wang & Tsinghua University & \href{mailto:anyuwang@tsinghua.edu.cn}{\texttt{anyuwang@tsinghua.edu.cn}}\\
\textbf{Xiaoyun Wang}\textsuperscript{\,\corrmail} & Tsinghua University & \href{mailto:xiaoyunwang@tsinghua.edu.cn}{\texttt{xiaoyunwang@tsinghua.edu.cn}}
\end{tabular}%
}{\corrmail\enspace Corresponding author.}

\end{document}